\documentclass[11pt]{article}
\usepackage{setspace,braket}
\usepackage{amsmath, amssymb, bm, yfonts}
\usepackage{jheppub}
\def\K{{\mathcal K}}

\def\be{\begin{equation}}
\def\ee{\end{equation}}
\def\lp{\ell_P}
\def\R{{\mathcal R}}
\def\Q{{\mathcal Q}}

\def\W{{\mathcal W}}

\def\K{{\mathcal K}}
\def\O{{\Omega}}
\def\B{{\mathcal B}}
\def\A{{\mathcal A}}
\def\C{{\mathcal C}}
\def\W{{\mathcal W}}

\def\be{\begin{equation}}
\def\ee{\end{equation}}
\def\lp{\ell_P}

\def\a{\alpha}
\def\b{\beta}

\def\r{\rho}
\def\l{\lambda}
\def\m{\mu}\def\n{\nu}\def\s{\sigma}\def\l{\lambda}

\def\bg{\bar{g}}
\def\beq{\begin{eqnarray}}\def\eeq{\end{eqnarray}}
\def\ba#1\ea{\begin{align}#1\end{align}}
\def\bg#1\eg{\begin{gather}#1\end{gather}}
\def\bm#1\em{\begin{multline}#1\end{multline}}
\def\bmd#1\emd{\begin{multlined}#1\end{multlined}}

\def\a{\alpha}
\def\b{\beta}

\def\ve{\varepsilon}

\def\l{\lambda}

\def\m{\mu}
\def\n{\nu}

\def\r{\rho}
\def\s{\sigma}

\def\la{\label}

\def\er{\eqref}

\def\fr{\frac}

\def\pa{\partial}
\def\td{\tilde}

\def\eq{\equiv}
\def\nn{\nonumber}

\def\lt{\left}
\def\rt{\right}
\def\({\left(}
\def\){\right)}
\def\[{\left[}
\def\]{\right]}

\def\Tr{{\rm Tr}}

\def\zb{{\bar z}}

\def\Area{{\rm Area}}

\newcommand{\equ}[1]{Eq.~(\ref{#1})}

\begin{document}
\title{On Entanglement Entropy Functionals in Higher-Derivative Gravity Theories}
\author[a]{Arpan Bhattacharyya}
\author[a,b]{Menika Sharma}
\affiliation[a]{Centre for High Energy Physics, Indian Institute of Science, C.V. Raman Avenue, Bangalore 560012, India}
\affiliation[b]{Harish-Chandra Research Institute, Chhatnag Road, Jhusi, Allahabad 211019, India}
\emailAdd{arpan@cts.iisc.ernet.in, menikasharma@hri.res.in}
\begin{abstract}{In arXiv:1310.5713~\cite{dong} and arXiv:1310.6659~\cite{camps} a formula was proposed as the entanglement entropy functional for a general higher-derivative theory of gravity, whose lagrangian consists of terms containing contractions of the Riemann tensor. In this paper, we carry out some tests of this proposal. First, we find the surface equation of motion for general four-derivative gravity theory by minimizing the holographic entanglement entropy functional resulting from this proposed formula. Then we calculate the surface equation for the same theory using the generalized gravitational entropy method of arXiv:1304.4926~\cite{maldacena}. We find that the two do not match in their entirety. We also construct the holographic entropy functional for quasi-topological gravity, which is a six-derivative gravity theory. We find that this functional gives the correct universal terms. However, as in the four-derivative case, the generalized gravitational entropy method applied to this theory does not give exactly the surface equation of motion coming from minimizing the entropy functional. }
\end{abstract}
\maketitle

\section{Introduction}

In the context of AdS/CFT, the entanglement entropy\footnote{There exists a huge literature on entanglement entropy. For background and interesting applications see  \cite{calabrese, myersme,  mukund, bum1, quantum,takatemp, New, bbkss, erd, AB}.} for a boundary field theory which is dual to Einstein gravity can be calculated using the well-known Ryu-Takayanagi proposal \cite{ryu,rev}. This proposal states that the entanglement entropy  $S_{EE}$ of any region on the boundary of AdS can be calculated by evaluating the area of a minimal surface in the bulk which is homologous to this boundary region:
\be
	S_{EE} = \fr{\Area}{4G_N} \,.
\label{arealaw}
\ee
Building upon earlier attempts~\cite{chm, head,  fursaev, fur1}, this proposal was recently proved in Ref.~\cite{maldacena}, for a general entangling surface.

The entanglement entropy formula in \equ{arealaw} is of the same form as the formula for calculating the entropy of a black hole. In the black hole case, there exists a simple generalization of this area law for calculating the entropy of a black hole in any general higher-derivative gravity theory, known as the Wald entropy \cite{wald,Iyer, Kang}.
It is natural to ask then if one can generalize the Ryu-Takayanagi prescription to higher-derivative gravity theories by simply replacing the RHS of \equ{arealaw} with the Wald entropy. However, this is known not to be the case \cite{higher,higher2}. 

Recently, a general formula for calculating the holographic entanglement entropy (HEE) in higher-derivative gravity theories was proposed in Refs.~\cite{dong,camps}. It was also conjectured that the minimal entangling surface can be determined by interpreting this formula as the entropy functional for the higher derivative gravity theory and extremizing it. At present there exists no general proof of this proposal. In this paper, we will carry out various tests to determine the validity of this conjecture. 

We will first work with general four-derivative theory. The conjectured form of the holographic entropy functional for general R$^2$ theory first appeared in Ref.~\cite{solo}. The formula of Refs.~\cite{dong, camps} also reduces to this functional for general R$^2$ theory. For the purpose of this paper, we will refer to this functional as the FPS (Fursaev-Patrushev-Solodukhin) functional after the authors of the paper where it was first proposed.
In Ref.~\cite{abs} it was shown that this entropy functional leads to the expected universal terms in the entanglement entropy for cylindrical and spherical entangling surfaces, so the FPS functional passes this basic first test. The obvious next step is to determine whether the surface equation of motion derived from extremizing this functional is the same as that derived using the generalized gravitational entropy method (which we will refer to as the LM method) of Ref.~\cite{maldacena}. 

General R$^2$ theory depends on three parameters: $\l_1,\l_2$ and $\l_3$. Gauss-Bonnet gravity is a special point in this parameter space \cite{gbholo} and the FPS functional reduces to the Jacobson-Myers functional at this point. For Gauss-Bonnet gravity, the question whether the surface equation of motion one gets from the Jacobson-Myers functional matches with the surface equation of motion derived using the LM method was addressed in Refs.~\cite{aks,abs,Chen}. We will look at the Gauss-Bonnet case again in this paper to emphasize several interesting points for this theory. For this theory, the surface equation of motion that one gets from the Jacobson-Myers functional matches with what one gets from the LM method, provided that terms cubic in the extrinsic curvature are suppressed. In this paper, we will find that for general R$^2$ theory using a procedure similar to the Gauss-Bonnet case leads to a match in the leading-order terms on both sides, where we designate terms cubic in the extrinsic curvature as sub-leading. However, as we will show, in the case of R$^2$ theory, the LM method also yields an extra condition that cannot be satisfied at arbitrary points of the parameter space. The conclusion is, therefore, that for a general R$^2$ theory the conditions that follow from the LM method do not correspond exactly to the surface equation of motion derived from the FPS functional.  

An alternative method to demonstrate that the FPS functional is the correct entropy functional for R$^2$ theory is to show that it can be interpreted as the action of a cosmic brane. This method was employed in Ref.~\cite{dong}, where it was referred to as the cosmic brane method. In this paper, we will re-examine this procedure for R$^2$ theory and show that the result we get is consistent with what we get using the LM method.

What happens when we go to a six-derivative gravity theory? In this case, we consider quasi-topological gravity \cite{quasi} which is again a special point in the parameter space of R$^3$ theories. We first construct the entropy functional for quasi-topological gravity using the formula proposed in Ref.~\cite{dong, camps}. We then show that this functional reproduces the expected universal terms for this theory for the cylindrical and spherical entangling surfaces. This is in agreement with the result of Ref.~\cite{miao} that the entropy functional proposed in Refs.~\cite{dong,camps} leads to the correct universal terms for a general higher-derivative theory. We also find the minimal surface condition for this theory using the LM method and show that it deviates from what is expected from the HEE functional. 

Our paper is organized as follows. In Sec.~(\ref{dongformula}) we review the general entropy functional proposed in \cite{dong, camps}. Our main focus in this paper is general four-derivative gravity theory, for which the entropy functional is the FPS functional. In Sec.~(\ref{R2}) we find the surface equation of motion for R$^2$ theory by extremizing the FPS functional on the codimension-2 surface. We then compare it with what we obtain using LM prescription. We also make some remarks on the Gauss-Bonnet case. We then investigate the cosmic-brane method of Ref.~\cite{dong}. In Sec.~(\ref{Quasi}), we repeat our analysis for quasi-topological gravity. Lastly, in Sec.~(\ref{discussion}) we summarize our findings and discuss their implications.


\section{Proposed entropy functional for general theories of gravity \label{dongformula}}

In this section we will review the general entropy formula proposed in \cite{dong,camps}. First we summarize the argument leading up to this proposal, following \cite{dong}. For details the reader is referred to \cite{dong, camps, maldacena, solo}. Some applications of this entropy formula are in \cite{extra}.

In field theory, the entanglement entropy $S_{EE}=-\Tr[\r\log\r]$ can be calculated as the $n\to1$ limit of the R\'enyi entropy.
The R\'enyi entropy in turn can be computed as
\be\la{renyi}
S_n= -\frac{1}{n-1}(\log Z_n-n\log Z_1) \,.
\ee
Here $Z_n$ is the partition function of the field theory on the manifold $M_n$ which is the $n$-fold cover of the  original spacetime manifold $M_1$. In the holographic dual theory one can construct a suitable bulk solution $B_n$ with boundary $M_n$. The manifold $M_n$ at integer $n$ has a $Z_n$ symmetry that cyclically permutes the $n$ replicas. In \cite{maldacena} it was proposed that this replica symmetry extends to the bulk $B_n$. Orbifolding the bulk by this symmetry results in a space $\hat B_n = B_n / Z_n \,,$ that is regular except at the fixed points of the $Z_n$ action. The fixed points form a codimension 2 surface with a conical defect in the bulk. This is the surface that is ultimately identified with the minimal entangling surface in the $n\rightarrow 1$ limit. 
Further, one can use gauge-gravity duality \cite{AdS} to identify the field theory partition function on $M_n$ with the on-shell bulk action on $B_n$ in the large-$N$ limit
\be\la{adscft}
Z_n\eq Z[M_n]= e^{-S[B_n]}\,.
\ee                                                       
It is now straightforward to calculate the entanglement entropy. By construction, one can identify
\be\label{sorb}
S[B_n] = n S[\hat B_n]
\ee
at integer $n$, where $S[\hat B_n]$ is the classical action for the bulk configuration $\hat B_n$ not including any contribution from the conical defect.  By analytically continuing $\hat B_n$ to non-integer $n$, \equ{sorb} can be used to define $S[B_n]$.  Using Eqs.~\er{renyi} and \er{adscft} and expanding around $n=1$, one gets
\be\label{snbh}
S_{EE} = \lim_{n\to1} \frac{n}{n-1}\left(S[\hat B_n]-S[ B_1]\right)  = \left . \partial_\epsilon S[\hat B_n]\right|_{\epsilon=0} 
\ee
where $\epsilon\equiv n-1$.
The quantity $S[\hat B_n]$ can be calculated for the bulk theory by writing the bulk metric locally around the surface in gaussian normal coordinates and introducing a conical defect. It can be shown \cite{maldacena, dong} that $\partial_\epsilon S[\hat B_n]$ gets a contribution entirely from the tip of the cone. To compute it, therefore, one employs a metric regularized at the tip of the cone.  

This calculation is similar to that employed in Ref.~\cite{solod} for calculating the Wald entropy from a regularized cone metric. Indeed, evaluating $ S[\hat B_n]$ for a bulk theory with the cone metric to linear order in $\epsilon$ and using \equ{snbh} will result in two terms. The first is $S_{\rm Wald}$: the Wald entropy for the theory. However, as was noted in \cite{solo}, there is a second way for a linear contribution to arise. A term in the bulk lagrangian that is of order $\epsilon^2$ can get enhanced to order $\epsilon$ after integrating over the transverse directions. Following \cite{dong}, we label the contribution of such terms as $S_{\rm Anomaly}$. At this point, the calculation of the form of $S_{EE}$ is basically finished. \equ{snbh} can be used to find the entanglement entropy for any higher-derivative theory, including ones whose lagrangians involve derivatives of the Riemann tensor.  However, for a general higher-derivative theory it can be computationally difficult to compute $S_{\rm Anomaly}$ directly using \equ{snbh}.

In \cite{dong,camps} a simpler prescription for calculating the holographic entanglement entropy for higher-derivative theories of gravity in $d+1$ dimensions, for which the lagrangian ($\mathcal{L}$) contains only contractions of the Riemann tensor, was given. The formula is:
\be\la{eei}
S_{EE}= 2\pi\int d^d y \sqrt{h} \lt\{ \fr{\pa \mathcal{L}}{\pa R_{z\zb z\zb}} + \sum_\a \(\fr{\pa^2 \mathcal{L}}{\pa R_{zizj} \pa R_{\zb k\zb l}}\)_\a \fr{8 \K_{zij} \K_{\zb kl}}{q_\a+1} \rt\} \,.
\ee
The notation used in the above equation and also in the rest of the paper is as follows: We use Greek Letters $\mu,\,\nu,\,\rho,\,\sigma,\, \cdots$ to label the bulk indices. We use the Latin letters $a,b, ...... m, n$ to label the indices of the codimension 2 surface, while reserving the letters $p,q,r,s$ to denote the indices of the transverse directions. In these directions, we use the complex coordinates $z$ and $\bar z$. The bulk metric is denoted by $g_{\mu\nu}$.The metric on the codimension-2 entangling surface is denoted by $h_{ij}$ while the surface itself is denoted by $\Sigma$. The bulk Riemann tensor is denoted by $R_{\mu \nu\rho\sigma}$ while the intrinsic Riemann tensor of the surface is denoted by $\R_{ikjl}$. The extrinsic curvatures of the surface are denoted by $\K_{rij}$, where the first index labels the extrinsic curvature in the transverse directions.  We follow the curvature conventions in Ref.~\cite{Waldb}.

The first term in \equ{eei} is the Wald entropy. The second term is the correction to the Wald entropy and is evaluated in the following way: The second derivative of the lagrangian $\mathcal{L}$ is a polynomial in components of the Riemann tensor. We expand the components $R_{pqij}, R_{piqj}$ and $R_{ikjl}$ using
\ba
R_{pqij} &= \td R_{pqij} +  \K_{pjk} \K_{qi}^{k} - \K_{pik} \K_{qj}^{k} \,,\nn\\
R_{piqj} &= \td R_{piqj} +  \K_{pjk} \K_{qi}^{k} - \Q_{pqij} \,,\nn\\
R_{ikjl} &= \R_{ikjl} + \K_{pil} \K_{pjk} - \K_{pij} \K_{pkl} \,.
\la{rexpi}
\ea
Here, $\Q_{pqij} \eq \frac{1}{2}\pa_p \pa_q g_{ij}|_\Sigma$.  $\td R_{pqij}$ and $\td R_{piqj}$ can also be defined in terms of metric variables, but the exact definition is not needed here.  The variable $\a$ is used to label the terms in the expansion.  For each term labelled by $\a$, $q_\a$ is defined as the number of $\Q_{zzij}$ and $Q_{\zb\zb ij}$, plus one half of the number of $\K_{pij}$, $R_{pqri}$, and $R_{pijk}$. The final step is to sum over $\a$ with weights $1/(1+q_\a)$.  The quantities $\td R_{pqij}$, $\td R_{piqj}$, and $\R_{ikjl}$ can then be eliminated using \equ{rexpi}, resulting in an expression involving only components of $R_{\m\n\r\s}$, $\K_{pij}$ and $\Q_{pqij}$. 

To yield the entanglement entropy, the formula in \equ{eei} should be evaluated on the minimal entangling surface. This surface is supposed to be determined following the LM method. Refs.~\cite{dong,camps} also contain the proposal that the minimal surface can be determined by extremizing $S_{EE}$ as given in \equ{eei} ---  $S_{EE}$ therefore being the entanglement entropy functional for a general theory of gravity.


\section{Test of the entropy functional for R$^2$ theory \label{R2}}
In this section we consider general $R^2$ theory in five dimensions. The lagrangian for this theory is
\be
\mathcal{L}=\mathcal{L}_1+\mathcal{L}_2\,,
\ee
where 
\be
\mathcal{L}_1=R+\frac{12}{L^2} \label{Ei} 
\ee
is the usual Einstein-Hilbert lagrangian with the cosmological constant appropriate for five-dimensional AdS space and 
\be 
\mathcal{L}_2= \frac{L^2}{2}\(\lambda_1 R_{\alpha\beta\gamma\delta}R^{\alpha\beta\gamma\delta}+\lambda_2 R_{\alpha\beta}R^{\alpha\beta}+\lambda_3 R^2 \)
\label{Lr}
\ee  
is the R$^2$ lagrangian. 

The proposed entropy functional for this theory is
\be 
  S_{\rm EE,\, R^2}= S_{\rm Wald, \,R^2} + S_{\rm Anomaly, \,R^2}\,,
  \label{area}
\ee
where 
\begin{align}
 S_{\rm Wald,\, R^2} &= \frac{2\pi}{\lp^{3}}\int d^{3}x \sqrt{h} \big\{1+\tfrac{L^{2}}{{2}} \left(2\lambda_{3}R+\lambda_{2} R_{\mu\nu}n^{\nu}_{r}n^{r\mu}+ 2\lambda_{1} R_{\mu\nu\rho\sigma}n^{\mu}_{r}n^{\nu}_{s}n^{r\rho}n^{s\sigma}\right) \big\}\,,
\label{Wald}   \\ \textrm{ and~~~ }
 S_{\rm Anomaly, \,R^2}&=\frac{2\pi}{\lp^{3}}\int d^{3}x \sqrt{h} \big\{ \tfrac{L^{2}}{{2}} \big(-\textstyle{\frac{1}{2}} \lambda_{2} \K_{r}\K^{r}- 2\lambda_{1}\K_{sij}\K^{sij} \big)\big\}\,.
\label{Extra} 
\end{align}
As mentioned earlier, this entropy functional leads to the correct universal terms. In this section, we will further test this entropy functional by determining whether the surface equation of motion one gets from extremizing this functional is the same as the surface equation of motion one gets following the LM method. In Sec.~(\ref{variation}), we extremize the functional for R$^2$ theory. In this particular section, we will first find the surface equation of motion for this functional in a general spacetime background.  However, the Ryu-Takayanagi proposal and its generalizations are most precisely formulated in the AdS/CFT context, so eventually we will specialize to the AdS background. In Sec.~(\ref{LMmethod}) we find the surface equation of motion using the LM method. In this case, we will always assume that the bulk is AdS space. Since a variation of the LM method -- called the cosmic-brane method -- was used in Ref.~\cite{dong} to formulate a proof that the FPS functional is the correct entropy functional for R$^2$ theory, we also investigate this method in Sec.~(\ref{deltas}). 

\subsection{Minimal surface condition from the entropy functional \label{variation}}

To extremize the functional in \equ{area}, we follow the methods of Refs.~\cite{diffgeo, Stringcurv1, Stringcurv2}. We denote the surface we are going to extremize w.r.t the action in \equ{area} by $\Sigma$. The induced metric on $\Sigma$ is 
\be
  h_{ij} = e^{\mu}_{i} e^{\nu}_{j} g_{\mu\nu},
\ee
where $g_{\mu\nu}$ is the bulk metric and $e^{\mu}_{i} \equiv {\partial}_{i}X^{\mu}$ are the basis vectors tangent to the surface $\Sigma$, $X^{\mu}$ being the bulk coordinates. On the surface $\Sigma$, the $g_{ir}$ component of the bulk metric vanishes. The two normals to the surface are denoted by $n^{\mu}_{r}$ where $r=1,2$ are the transverse directions. The metric tensor in the tangent space spanned by the normal vectors (the metric tensor of the normal bundle of the sub-manifold $\Sigma$) is the Kronecker delta:
\be
	\delta_{rs}  = \epsilon \, n_{r}^{\mu} n^{\nu}_s g_{\mu\nu}
\ee
We work in Euclidean signature and set $\epsilon=+1$. We use the inverse metric $\delta^{rs}$, to raise indices in the normal directions: $n^{r \mu } =\delta^{rs} n^{\mu}_{s}$. Note that, repeated $s$ indices always imply summation over the transverse directions: $n^{\mu}_s n^{\nu s} \equiv n^{\mu}_1 n^{\nu}_1 + n^{\mu}_2 n^{\nu}_2$.
In this notation, the completeness relation relating $g^{\mu\nu}$, the inverse of the bulk metric, to $h^{ij}$, the inverse of the induced metric, is
\be
  g^{\mu\nu} = h^{ij}e^{\mu}_{i}e^{\nu}_{j} +  n^{\mu}_s n^{\nu s}. 
  \label{completeness}
\ee

The Gauss and Weingarten equations are 
\begin{align} \label{GaussW}
\nabla_{i} e^{\mu}_{j}&=\partial_{i} e^{\mu}_{j} + {\hat{\Gamma}}^{\mu}_{\nu\rho}e^{\rho}_{i} e^{\nu}_{j} -\Gamma_{ij}^{k}e^{\mu}_{k}= -\K^{r}_{ij} n^{\mu}_{r} \nn \\
\nabla_{i} n^{\mu}_{s}&= \partial_{i} n^{\mu}_{s} +{\hat{\Gamma}}^{\mu}_{\rho\nu} e^{\rho}_{i} n^{\nu}_s-\Gamma_{si}^{r}n_{r}^{\mu} =\K_{ s i}^{j} e^{\mu}_{j}\,.
\end{align}
Here, $\nabla$ is the Van der Waerden-Bortolotti covariant derivative~\cite{diffgeo} which acts on a general tensor $T^{s\cdots r}_{i\cdots j}$ as
\be
   \nabla_k T^{s\cdots r}_{i\cdots j} = \tilde{\nabla}_k T^{s\cdots r}_{i\cdots j} + {\Gamma}^{s}_{p k} T^{p\cdots r}_{i\cdots j} + \cdots +  {\Gamma}^{r}_{p k} T^{s\cdots p}_{i\cdots j}  \,,
   \label{defnabla}
\ee
where $\tilde{\nabla}$ is the usual covariant derivative associated with the surface Christoffel. This Christoffel is related to the bulk Christoffel ${\hat{\Gamma}}^{\mu}_{\sigma \nu}$ as 
\be
	{\Gamma}^{i}_{j k} = (\partial_j e^{\mu}_k  + {\hat{\Gamma}}^{\mu}_{\sigma \nu} e^{\sigma}_j e^{\nu}_k ) e^{i}_{\mu}\,.
\ee
The Chrisoffel ${\Gamma}^{r}_{i s}$ is the Christoffel induced in the normal bundle. It is related to the bulk Christoffel as
\be
	{\Gamma}^{r}_{i s} = (\partial_i n^{\mu}_s  + {\hat{\Gamma}}^{\mu}_{\sigma \nu} e^{\sigma}_i n^{\nu}_s) n^r_{\mu}\,.
	\label{defCA}
\ee
This Christoffel can be interpreted geometrically as the freedom to perform rotations of the normal frame. It is, therefore, equivalent to a gauge field $\A_{k}$, commonly referred to as a twist potential. This field is defined as:
\be
  \A_{k} \equiv \frac{1}{2} \varepsilon^{rs} \partial_r g_{ks}\,,\textrm{ so that } {\Gamma}^{s}_{j r} =\delta^{ps}\varepsilon_{rp} \A_{j} \,,
  \label{defA}
\ee
where $\varepsilon_{rs}$ is the Levi-Civita symbol. 

The Gauss identity relating the bulk Riemann with all indices projected in the tangential directions with the surface Riemann is
\be \label{GaussRicci}
		R_{\mu\nu\rho\sigma}e^{\mu}_{k}e^{\nu}_{i}e^{\rho}_{l}e^{\sigma}_{j}= \R_{kilj}-\K^{r}_{kl} \K_{rij} + \K^{r}_{ik}\K_{rjl}\,.
\ee
The Codazzi-Mainardi relation is
\be \label{codazzi}
	 \nabla_k \K_{rij}  -  \nabla_i \K_{rkj}  =  R_{\mu\nu\rho\sigma}e^{\mu}_{k}e^{\nu}_{i}e^{\sigma}_{j}n^{\rho}_{r}\,.
\ee
From \equ{GaussRicci} we get the Gauss-Codazzi identity
\be
  \R = R - 2 R_{\mu\nu}n^{\nu r}n^{\mu}_{r} + R_{\mu\nu\rho\sigma}n^{\mu r}n^{\nu s}n^{\rho}_{r}n^{\sigma}_{s} + \K_r\K^r - \K^s_{ij}\K_s^{ij}\,.
  \label{GC}
\ee

We now consider an infinitesimal variation of the surface $\Sigma$ given by $X^{\mu} \longrightarrow  X^{\mu} +\delta X^{\mu}$. The change $\delta X^{\mu}$ is
\be
  \delta X^{\mu} = {\xi}^r n^{\mu}_r +  {\xi}^i e^{\mu}_i\,.
  \label{fullvariation}
\ee
where ${\xi}^r$ and ${\xi}^i$ are small parameters. 
For deriving the equation describing the minimal surface we are only concerned with the variation in the normal direction, since the tangential variation leads to a constraint equation. The variation then reduces to
\be
  \delta X^{\mu} = {\xi}^r n_{r}^{\mu}\,,
\ee 

The variation $\delta X^{\mu}$ in the surface will induce a variation in the basis vectors $e_{i}^{\mu}$. This can be computed by finding the basis vectors at $X^{\mu} +\delta X^{\mu}$ and parallel transporting them back to $X^{\mu} $. Taking the difference between the parallel-transported quantity and the original basis vector at the coordinate $X^{\mu}$, using the identities in \equ{GaussW} and then restricting to normal variation results in 
\be
  \delta e_{i}^{\mu} =  n^{\mu}_{s} {\nabla}_i \xi^s  + e_{j}^{\mu} \K^{j}_{si}  \xi^s\,.
  \label{tangentvar}
\ee
The details of this calculation are in Ref.~\cite{diffgeo}.
As stated in \equ{defnabla}, the covariant derivative $\nabla$ acts on $\xi_s$ as
\be
	\nabla_i \xi^s = \partial_i \xi^s +  {\Gamma}^{s}_{i r} \xi^{r}\,.
\ee

The variation in any other tensor quantity can be calculated in a similar way, by parallel transporting the quantity at the new coordinate back to the old coordinate and taking the difference. This gives the variation in the bulk metric as zero. We write down the result for other variations. For details the reader is referred to \cite{diffgeo}.
The variation of the induced metric is 
\begin{align}
  \delta h_{ij} &=  2 {\xi}^r \K_{rij}\,, \nonumber \\
  \delta \sqrt{h} &=  {\xi}^r \sqrt{h}  \K_r\,.
  \label{metricvar}
\end{align}

The variation of the extrinsic curvature is 
\begin{align}
  \delta \K^s_{ij} &= (- {\nabla}_{i} {\nabla}_{j} {\delta}^{s}_{r} + \K^{s}_{i k} \K^{k}_{rj} + R_{\mu\nu\rho\sigma}n^{s\mu}n^{\sigma}_r e^{\rho}_{i} e^{\nu}_{j} ) {\xi}^{r}\,,\nonumber \\
\delta \K^s &= ( - {\nabla}^{i} \nabla_{i} {\delta}^{s}_{r} - \K^{s}_{i k} \K^{ki}_{r} + h^{ij}R_{\mu\nu\rho\sigma}n^{s\mu}n^{\sigma}_r e^{\rho}_{i} e^{\nu}_{j} ) {\xi}^{r}\,.
\label{Kvar}
\end{align}
The covariant derivatives $\nabla$ act all the way to the right.

Using these variations we can now compute the change in the action. For this we first rewrite the $R_{\mu\nu}n^{\nu}_{r}n^{r\mu}$ and $R_{\mu\nu\rho\sigma}n^{\mu}_{r}n^{\nu}_{s}n^{r\rho}n^{s\sigma}$ terms in the action given in Eq.~(\ref{Wald}) as
\begin{align}
  R_{\mu\nu}n^{\nu r}n^{\mu}_{r} &= R - h^{ij} R_{\mu\nu}e^{\nu}_{i}e^{\mu}_{j} \, \nonumber \\
  R_{\mu\nu\rho\sigma}n^{\mu r}n^{\nu s}n^{\rho}_{r}n^{\sigma}_{s} &= R - 2 h^{ij} R_{\mu\nu}e^{\nu}_{i}e^{\mu}_{j} + h^{ik}h^{jl}R_{\mu\nu\rho\sigma}e^{\mu}_{i}e^{\nu}_{j}e^{\rho}_{k}e^{\sigma}_{l}
\end{align}
using the completeness relation in Eq.~(\ref{completeness}). The variation of a term such as $h^{ij} R_{\mu\nu}e^{\nu}_{i}e^{\mu}_{j}$ is given by
\be
	\delta ( h^{ij} R_{\mu\nu}e^{\nu}_{i}e^{\mu}_{j} ) = ( \delta h^{ij} )R_{\mu\nu}e^{\nu}_{i}e^{\mu}_{j} + 2  h^{ij} R_{\mu\nu} \delta (e^{\nu}_{i}) e^{\mu}_{j} + h^{ij} \delta (R_{\mu\nu}) e^{\nu}_{i}e^{\mu}_{j}
\ee
The first two variations can be computed using Eqs.~(\ref{metricvar}) and (\ref{tangentvar}) respectively.  
For evaluating the last term we need the variation of the bulk Ricci Tensor\footnote{We thank Joan Camps for pointing out that such terms will contribute to the total variation.} which is given by 
\be
	\delta (R_{\mu\nu}) =  n^{\sigma}_{r} {\hat{\nabla}_\sigma}R_{\mu\nu}  {\xi}^{r}\,.
\ee
The variation in the bulk Ricci scalar and Riemann tensor take a similar form. All these variations are under the integral sign in \equ{Wald} and we  perform a integration by parts where needed, discarding the term that is a total variation. Then 
using the variations given above we obtain:
\begin{alignat}{2} 
  \delta(\sqrt{h} R) &= && \sqrt{h}~ \K_{s} R \,{\xi}^s + n^{\mu}_{s} \sqrt{h} \hat \nabla_{\mu}R\,{\xi}^{s}\,, \nn \\
  \delta(\sqrt{h} R_{\mu\nu}n^{\nu r}n^{\mu}_{r}) &= && \sqrt{h}~ \K_{s} R_{\mu\nu}n^{\nu r}n^{\mu}_{r} {\xi}^s + 2 \sqrt{h} \,\nabla^{i}(R_{\mu\nu}n^{\nu}_s e^{\mu}_{i}) {\xi}^s~-\nonumber\\
	  &\phantom{=}&&\sqrt{h} \,  n^{\sigma}_{s}h^{ij} e^{\mu}_{i}e^{\nu}_{j}\hat \nabla_{\sigma}R_{\mu\nu}{\xi}^{s} + n^{\mu}_{s} \sqrt{h} \,\hat \nabla_{\mu}R\,{\xi}^{s}\,, \nn \\
	\delta(\sqrt{h} R_{\mu\nu\rho\sigma}n^{\mu r}n^{\nu s}n^{\rho}_{r}n^{\sigma}_{s}) &=  && \sqrt{h}~ \K_{s} R_{\mu\nu\rho\sigma}n^{\mu r}n^{\nu q}n^{\rho}_{r}n^{\sigma}_{q} {\xi}^s-
	4 \sqrt{h}  \,\nabla^{i}(R_{\mu\nu\rho\sigma} n^{\mu}_s e^{\nu}_{j}e^{\rho}_{i}e^{\sigma}_{k}h^{jk} ){\xi}^s + \nonumber\\
	  &\phantom{=}&&4 \sqrt{h}\,\nabla^{i}( R_{\mu\nu}n^{\nu}_s e^{\mu}_{i}) {\xi}^s+\sqrt{h}  h^{ik}h^{jl}e^{\mu}_{i}e^{\nu}_{j}e^{\rho}_{k}e^{\sigma}_{l} n^{\alpha}_{s} \hat \nabla_{\alpha} R_{\mu\nu\rho\sigma}{\xi}^{s}  ~ -\nonumber\\
	  &\phantom{=}&& 2\,n^{\nu}_{s} h^{ij} e^{\mu}_{i}e^{\rho}_{j}\hat \nabla_{\nu}R_{\mu\rho}{\xi}^{s}+\sqrt{h} n^{\mu}_{s}\hat \nabla_{\mu}R{\xi}^{s} \,.
	 \label{varR}
\end{alignat}
Similarly the variations for the terms present in the action in Eq.~(\ref{Extra}) are
\begin{alignat}{2}
  \delta(\sqrt{h}\K^s\K_{s}) &= &-& 2 \sqrt{h}\nabla_i\nabla^i \K_{r} {\xi}^r +  \sqrt{h} \K_{r} \K^s \K_{s}  {\xi}^r - 2 \sqrt{h} \K^{s} \K_{sij}\K_r^{ij}  {\xi}^r - \nonumber\\
    &\phantom{=} &&2 \sqrt{h} \K^{s} R_{\mu\nu\rho\sigma} h^{ij} n_r^{\mu} e^{\nu}_{i}n_s^{\rho}e^{\sigma}_{j}{\xi}^r \,, \nonumber\\ 
  \delta(\sqrt{h}\K_{sij}\K^{sij}) &=&-&  2 \sqrt{h} \nabla_i\nabla_j \K_r^{ij} {\xi}^r +  \sqrt{h}\K_{r} \K_{sij}\K^{sij}  {\xi}^r - 2\sqrt{h} \K_{sij}\K^{si}_{k} \K_{r}^{kj}{\xi}^r - \nonumber\\
    &\phantom{=}&&  2 \sqrt{h}\K^{sij} R_{\mu\nu\rho\sigma}n^{\mu}_r e^{\nu}_{i}n^{\rho}_s e^{\sigma}_{j}{\xi}^r \,.
  \label{varK2}
\end{alignat}
Adding these variations up with the appropriate factors will give us the equation for the minimal surface for the action in Eq.~(\ref{area}) in a general spacetime background.

As a check of these equations we now demonstrate that the above results lead to the correct surface equation of motion in the Gauss-Bonnet case. For Gauss-Bonnet the entropy functional for general R$^2$ theory reduces to the Jacobson-Myers functional 
\be
  S_{JM}=\frac{2\pi}{\lp^3} \int d^3 x \sqrt{h}\{(1+\lambda L^2 (R - 2 R_{\mu\nu}n^{\nu r}n^{\mu}_{r} + R_{\mu\nu\rho\sigma}n^{\mu r}n^{\nu s}n^{\rho}_{r}n^{\sigma}_{s} + \K_s\K^s - \K_{sij}\K^{sij})\}\,.
\label{JMbig}
\ee
This functional is valid in a general space-time background. Note that the integrand is equal to $\sqrt{h}(1+\lambda L^2 \R)$ on using the Gauss-Codazzi identity \equ{GC}. The surface equation of motion for this theory using this form of the functional is
~\cite{abs},
\be
   \K + \lambda L^2 (\R \K - 2\R^{ij}\K_{ij}) =0\,.
   \label{GBminimal}
\ee
We now find the surface equation of motion by directly varying \equ{JMbig}. Using the variation equations Eqs.~(\ref{varR}--\ref{varK2}) and simplifying using the identities in Eqs.~(\ref{GaussRicci}--\ref{codazzi}) we get 
\begin{alignat}{2} \label{GS2}
\sqrt{h}~\K_s\,{\xi}^s+\lambda L^2\Big[&\sqrt{h}~ \K_{s} \R \,{\xi}^s- 2\,\sqrt{h} \R_{jk}\K_s^{jk}\, {\xi}^s \nonumber\\\phantom{=}&+\sqrt{h}  h^{ik}h^{jl}e^{\mu}_{i}e^{\nu}_{j}e^{\rho}_{k}e^{\sigma}_{l} n^{\alpha}_{s} \hat \nabla_{\alpha} R_{\mu\nu\rho\sigma}{\xi}^{s} -
	2 \sqrt{h}\,\nabla^{i}( R_{\mu\nu\rho\sigma} n^{\mu}_s e^{\nu}_{j}e^{\rho}_{i}e^{\sigma}_{k}h^{kj} ){\xi}^s  \nonumber\\\phantom{=}&-2 \sqrt{h} \K^{r} R_{\mu\nu\rho\sigma} h^{ij} n_s^{\mu} e^{\nu}_{i}n_r^{\rho}e^{\sigma}_{j}{\xi}^s +2 \sqrt{h}\K^{rij} R_{\mu\nu\rho\sigma}n^{\mu}_s e^{\nu}_{i}n^{\rho}_r e^{\sigma}_{j}{\xi}^s  \nonumber\\\phantom{=}&+2 \sqrt{h} R_{\mu\nu\rho\sigma}e^{\mu}_{j}e^{\nu}_{i}e^{\rho}_{k}e^{\sigma}_{l}h^{il}\K_s^{jk} \, {\xi}^s\Big]\,.
\end{alignat}
The first three terms give precisely the equation of motion for Gauss-Bonnet theory. The rest of the terms add up to zero, as we show in the following.
We use the Bianchi identity on the $\hat \nabla_{\alpha} R_{\mu\nu\rho\sigma}$ factor of the fourth term giving
\be
\hat \nabla_{\alpha} R_{\mu\nu\rho\sigma}= -\hat \nabla_{\sigma} R_{\mu\nu\alpha\rho}-\hat\nabla_{\rho} R_{\mu\nu\sigma \alpha}
\ee
and then rewrite each of these terms as 
\be \label{leftovera}
  h^{ik}h^{jl}e^{\mu}_{i}e^{\nu}_{j}e^{\rho}_{k}e^{\sigma}_{l} n^{\alpha}_{s} \hat \nabla_{\sigma} R_{\mu\nu\alpha\rho}~=~ e^{\sigma}_{l} \hat \nabla_{\sigma}(h^{ik}h^{jl}e^{\mu}_{i}e^{\nu}_{j}e^{\rho}_{k} n^{\alpha}_{s} R_{\mu\nu\alpha\rho})- 
e^{\sigma}_{l} \hat \nabla_{\sigma}(h^{ik}h^{jl}e^{\mu}_{i}e^{\nu}_{j}e^{\rho}_{k} n^{\alpha}_{s}) R_{\mu\nu\alpha\rho}\,.
\ee
The expression in brackets in the first term of the R.H.S is a bulk scalar and therefore this term can be written as 
\begin{align}
 \partial_{l}(h^{ik}h^{jl}e^{\mu}_{i}e^{\nu}_{j}e^{\rho}_{k} n^{\alpha}_{s} R_{\mu\nu\alpha\rho}) =  -&\nabla^{i}( R_{\mu\nu\rho\sigma} n^{\mu}_s e^{\nu}_{j}e^{\rho}_{i}e^{\sigma}_{k}h^{jk} )+ \Gamma^{r}_{sl}h^{ik}h^{jl}e^{\mu}_{i}e^{\nu}_{j}e^{\rho}_{k} n^{\alpha}_{r} R_{\mu\nu\alpha\rho} + \nonumber \\&\Gamma _{jl}^{m}h^{ik}h^{jl}e^{\mu}_{i}e^{\nu}_{m}e^{\rho}_{k} n^{\alpha}_{s} R_{\mu\nu\alpha\rho}\,,
\end{align}
Inserting these expressions in \equ{GS2} after expanding the second term on the R.H.S of \equ{leftovera} and using the identities in \equ{GaussW} will lead to cancellation of all terms except for the terms in the first line of \equ{GS2}.

\subsubsection*{AdS background \label{ADSeom}}

We now specialize to the case of AdS background which is the background we will use while finding the equation of motion using the LM method. In AdS space the Riemann tensor is
\be
     R_{\mu\nu\rho\sigma} = -C(g_{\mu\rho}g_{\nu\sigma}-g_{\mu\sigma}g_{\nu\rho}) \,,
     \label{ADSRiemann}
\ee
where we have defined $C=f_{\infty}/{L}^2$. Here, $L$ is the length scale associated with the cosmological constant and is related to the AdS radius $\tilde L$ as $L =\tilde L \sqrt{f_{\infty}}$. The variable $f_{\infty}$ satisfies the following equation for $R^2$ theory
\be
1-f_\infty+\frac{1}{3}f_\infty^2 (\lambda_1+2\lambda_2+10\lambda_3)=0\,.
\ee

For ease of comparison with later results, we also rewrite the variation in $\sqrt{h}R$ given in Eq.~(\ref{varR}) using the Gauss-Codazzi relation \equ{GC}. The minimal surface equation is then
\begin{align}
   \K^{r} + L^2 \{&\lambda_3(\R \K^r - 2\R^{ij}\K^{r}_{ij} + 2\nabla^2 \K^{r} - 2 \nabla_i\nabla_j \K^{rij} -\nonumber\\& ~~~~ \K^{r} \tilde{\K}_2  + 2 \K^{r}_{ij} \K_{2}^{ij} +  \K^{r} \K_2 - 2 \K_3^{r}  - 18 C \K^{r} ) + \nonumber \\
   &\lambda_2( \tfrac{1}{2} \nabla^2 \K^{r}  - \tfrac{1}{4} \K^{r} \tilde{\K}_2 + \tfrac{1}{2} \K^{r}_{ij} \K_{2}^{ij} - \tfrac{11}{2} C \K^{r} ) + \nonumber \\
   &\lambda_1(2 \nabla_i\nabla_j \K^{rij} -  \K^{r} \K_2 + 2 \K_3^{r} - 4 C \K^r ) \} =0\,,
   \label{R2minimalo}
\end{align}
where we have defined $\K_2= \K_{sij}\K^{sij},\K_{2}^{ij} = \K_{s}\K^{sij}, \tilde{\K}_2 = \K_s \K^{s}$ and $\K_3^{r}= \K_{sij}\K^{si}_k \K^{rkj}$. Note that these are a set of two equations one for each of the extrinsic curvatures $\K^1,\K^2$.

In AdS space we can make a further simplification using \equ{codazzi}. The R.H.S of this equation disappears on using \equ{ADSRiemann}. We then get the identity $\nabla^k \nabla_k\K_{r}= \nabla^i \nabla^j \K_{rij} $ on taking a further covariant derivative of the L.H.S. As explained in Appendix~\ref{K2}, in the LM method for a time-independent metric, we can set $\K^{1}=\K^{2}=\K$. We, therefore, also drop the $r$ index and \equ{R2minimalo} simplifies to
\begin{align}
   \K + L^2 \{&\lambda_3(\R \K - 2\R^{ij}\K_{ij}  - \K^3 + 3 \K  \K_2 - 2 \K_3 - 18 C \K ) + \nonumber \\
   &\lambda_2( \tfrac{1}{2} \nabla^2 \K  - \tfrac{1}{4} \K^3 + \tfrac{1}{2} \K \K_{2} - \tfrac{11}{2} C \K ) + \nonumber \\
   &\lambda_1(2 \nabla^2 \K  -  \K \K_2 + 2 \K_3 - 4 C \K) \} =0\,.
   \label{R2minimal}
\end{align}
We have also verified this equation by determining the bulk extremal surfaces for different types of boundary entangling regions (sphere, cylinder and slab).  

For the Gauss-Bonnet case: $\lambda_1=\lambda, \lambda_2=-4\lambda$ and $\lambda_3=\lambda$, this equation reduces to the known result in \equ{GBminimal}. Note that terms cubic in the extrinsic curvature as well as the $C \K$ terms are not present in that equation. The Gauss-Bonnet case is special in this sense. No such simplification occurs if we set the value for Weyl$^2$ theory, $\lambda_1=\lambda,\lambda_2=-4\lambda/3$ and $\lambda_3=\lambda/6$:
\be
   \K + \frac{\lambda L^2}{6} (\R \K - 2\R^{ij}\K_{ij} + 8\nabla^2 \K  +   \K^3   -7 \K \K_2  +  10 \K_3  +2 C\K )  =0 \,.
   \label{Weyleom}
\ee
The $C \K$  term, in particular, stands out. If we trace the provenance of this term, it comes from terms of the form $\K_{s} R_{\mu\nu\rho\sigma}n^{\mu}_{r}n^{\nu}_{q}n^{\rho r}n^{\sigma q}$ and $\K^{ij}_{s} R_{\mu\nu\rho\sigma}n^{\mu}_{r}e^{\nu}_{i}n^{\rho}_{s}e^{\sigma}_{j}$ in Eqs.~(\ref{varR}) and (\ref{varK2}) --- terms that have components normal to the surface. Nevertheless, for AdS background these reduce to a term intrinsic to the surface. In fact, using the Gauss-Codazzi identity, Eq.~(\ref{GC}), we can rewrite this $C \K$  term as $\sim \K^3 + \R\K$. 

So far, we have only considered normal variations of the surface. Considering tangential variations leads to a constraint equation. For R$^2$ theory this constraint equation is indistinguishable from the condition in \equ{codazzi} which is the Codazzi-Mainardi relation.


\subsection{Minimal surface condition from the Lewkowycz-Maldacena method \label{LMmethod}}

We will now derive the surface equation for R$^2$ using the LM method. As already mentioned, the main idea of Ref.~\cite{maldacena} is that one can obtain the minimal surface condition by extending the replica trick to the bulk. 
The bulk will then have a $Z_n$ symmetry. Orbifolding by this symmetry will lead to a spacetime in which the fixed points form a codimension-2 surface with a conical deficit. In the $n \rightarrow 1$ limit this surface can be identified with the entangling surface. The metric of this surface can be parametrized in gaussian normal coordinates as follows: 
\begin{align}
ds^2 =  e^{2\rho(z,\bar z)} \{dz d\zb~ +~ &e^{2\rho(z,\bar z)} \O (\zb dz-z d\zb)^2 \} + (g_{ij} + \K_{rij} x^r + \Q_{rsij} x^r x^s ) dy^i dy^j ~+ \nonumber \\ 
&2 e^{2\rho(z,\bar z)} (\A_i + \B_{ri} x^r) (\zb dz-z d\zb) dy^i + \cdots \,.
\label{metric2}
\end{align}
Here $\rho(z,\bar z)= -\frac{\epsilon}{2} \ln( z\bar z)$ and $\epsilon=n-1$, while $x^1=z$ and $x^2=\bar{z}$.
This is the most general form of the metric upto terms second order in $z(\zb)$~\cite{solo,dong,camps, smolkin}. The $\cdots$ denote higher-order terms. As we will see later, for $R^2$ theory we also need to include third-order terms in the metric expansion.
The quantity $\K_{ij}$ in this metric is identified with the extrinsic curvature, while $\A_i$ is identified with the twist potential. Both of these are standard quantities that characterize the embedding of the surface in the bulk. The quantities $\O, \B_{ri}$ and $\Q$ in the second-order terms in the metric are not arbitrary, but can be written in terms of $\K_{rij},\A_i$ and the components of the curvature tensors.

The bulk equation of motion will now contain divergences arising out of the conical singularity of the form $\frac{\epsilon}{z},\frac{\epsilon}{z^2}$. However, the matter stress-energy tensor is expected to be finite. Therefore, we must set all divergences to zero. This condition fixes the location of the entangling surface. 

The bulk equation of motion for general four-derivative theory is
\cite{Sinha:2010pm}:
\be
R_{\alpha\beta}-\frac{1}{2}g_{\alpha\beta}R- \frac{6}{L^2}g_{\alpha\beta}-\frac{L^2}{2} H_{\alpha\beta}=0\,,
\label{bulkeom}
\ee
where
\begin{align}
  H_{\alpha\beta}~=~ &\lambda_1 \Big(\frac{1}{2}g_{\alpha\beta} R_{\delta \sigma \mu 
  \nu} R^{\delta\sigma \mu\nu}-2 R_{\alpha \sigma \delta \mu}{ R_\beta}{}^{\sigma \delta 
  \mu}-4 \hat \nabla^2  R_{\alpha \beta}+2\hat \nabla_\alpha \hat \nabla_\beta  R~+~\nonumber\\&~~~~~4  R_\alpha^\delta  R_{\beta \delta}+4 R^{\delta \sigma} R_{\delta \alpha \beta \sigma}\Big) +  \nonumber \\
  &\lambda_2\Big(\hat\nabla_{\alpha}\hat \nabla_{\beta} R+2 R^{\delta \sigma} R_{\delta \alpha \beta \sigma}  - \hat \nabla^2  R_{\alpha\beta}+\frac{1}{2} g_{\alpha\beta}  R_{\delta \sigma} R^{\delta\sigma}-\frac{1}{2} g_{\alpha\beta} \hat \nabla^2  R\Big) + \nonumber \\ 
  &\lambda_3\Big(-2  R  R_{\alpha\beta}+2 \hat \nabla_\alpha \hat \nabla_\beta  R +\frac{1}{2}g_{\alpha\beta} R^2 - 2 g_{\alpha\beta} \hat \nabla^2  R\Big) \,.
  \label{Hab}
\end{align}

\subsubsection{Gauss-Bonnet theory revisited}
Our eventual goal is to find the surface equation of motion for general R$^2$ theory, but it is illuminating to look at the Gauss-Bonnet case first. The Gauss-Bonnet case was addressed in Refs.~\cite{aks, abs, Chen} using a metric linear in $z(\zb)$. In this section, we will find the surface equation of motion for this theory using the metric in \equ{metric2}. 

We first show that the second-order metric in \equ{metric2} suffices for Gauss-Bonnet theory and inclusion of higher-order terms in this conical metric will not affect the surface equation of motion that we find for this theory from the LM method. 
The bulk equation of motion for Gauss-Bonnet theory can be obtained from \equ{bulkeom} by setting $\lambda_1=\lambda,\lambda_{2}=-4\lambda$ and $\lambda_{3}=\lambda$ giving:
\begin{align}
  H_{\alpha\beta}~=~&4  R_\alpha^\delta  R_{\beta \delta}-4 R^{\delta \sigma} R_{\delta \alpha \beta \sigma}-2  R  R_{\alpha\beta} -2 R_{\alpha \sigma \delta \mu}{ R_\beta}{}^{\sigma \delta 
  \mu} + \nn \\
& \tfrac{1}{2}g_{\alpha\beta} (R_{\delta \sigma \mu 
  \nu} R^{\delta\sigma \mu\nu}-4R_{\delta \sigma} R^{\delta\sigma}+R^2) \,.
 \label{HabGB}
\end{align}
The surface equation of motion is derived by finding the divergences in this equation that arise on using the conical metric in the limit $z=\zb\rightarrow 0$. Terms higher than second-order in the metric will not contribute to the curvature tensors to zeroeth-order in $z(\zb)$, although they might contribute at higher order. This is because the curvature tensors are of dimension $1/\rm{Length}^2$ while third-order terms in the metric will be of order $1/\rm{Length}^3$. The explicit values of the curvature tensors are listed in Appendix~(\ref{append}). These are calculated using a conical metric which is third-order in $z(\zb)$. Note also, that the curvature tensors contain at most divergences of order $1/z$. In the above equation of motion all terms are the product of two curvature tensors. Since each curvature tensor can only contribute at most a $1/z$ divergence and no third(or higher)-order term occurs at zeroeth order in any curvature tensor, third(and higher)-order  terms will be absent in the divergence equations.

By the same logic one can see that second-order terms will contribute to the divergence equations. However, in this case, cancellations between terms remove most of the second-order quantities, leaving only the quantities $\Q_{zzij}$ and $\Q_{\zb \zb ij}$ in the divergence equations.

In the $z=\zb\rightarrow 0$ limit $\K^1=\K^2$ as explained in Appendix~\ref{K2}, so we drop the index $r$ on $\K^r$.
The divergence in the $zz$ component from $H_{\a\b}$ term in the bulk eom is 
\begin{align}
~~~~~H_{zz} ~=~\frac{\epsilon }{ z}&\Big[\lambda (\R\K-2 \K^{ij} \R_{ij})\Big]+\frac{\epsilon}{z}\Big[e^{-2\rho(z,\bar{z})}\lambda\Big\{-\K^3+3\K\K_{2}-2\K_3\Big\}\Big] \,.
\label{zzcomponentgb}
\end{align}
Setting this divergence to zero should yield the condition for the extremal surface.
There is no divergence in the $z\zb$ component. The divergence in the $zi$ component is
\begin{align}
H_{zi}~= ~\frac{\epsilon }{z}&\Big[e^{-2\rho(z,\bar z)}\lambda\Big\{2\K \nabla_i \K -2\K\nabla_j\K^j_i-2\K^j_i\nabla_j \K+2\K_{ij}\nabla_k \K^{kj}~-~\nonumber\\&~~~~~~~~~~~~~~~2\K_{kj}\nabla_i \K^{kj}+2 \K_{jk}\nabla^j \K^{k}_i\Big\}\Big]\,.
\label{zicomponentgb}
\end{align}
This divergence is equivalent to the constraint equation one gets for the entropy functional (which doesn't have to be necessarily the Jacobson-Myers functional) using tangential variations of the surface and vanishes similarly by \equ{codazzi}.
Finally, the divergence in the $ij$ component is
\begin{align}
~~~~H_{ij}~=~\frac{4\epsilon }{z}&\Big[e^{-4\rho(z,\bar z)}\lambda\Big\{2\K_{ik}\K^{kl}\K_{lj}+h_{ij}\K\K_2-
\K_{ij}\K_2-h_{ij}\K_3-\K\K_{ik}\K^{k}_{j}~-~4 h_{ij}\K \Q_{zz}\nonumber\\&~~~~~~~~~~~~~~~+4 h_{ij}\K_{kl}\Q_{zz}^{kl}~-~
8 \K_{ki}\Q^{k}_{zzj}+
4 \K_{ij}\Q_{zz}~+~4\K\Q_{zzij}\Big\}\Big]~+\nonumber\\ \frac{2\epsilon^2 }{z^2}&\Big[e^{-4\rho(z,\bar z)}\lambda\Big\{2\K_{ij}\K-2\K_{ik}\K^{k}_{j}-h_{ij}\K^2+h_{ij}\K_2
\Big\}\Big]\,.\label{ijcomponentgb}
\end{align}
In the above equation we have set $\Q_{zzij}=\Q_{\zb \zb ij}$. Using the value of the $R_{zizj}$ component of the Riemann tensor from Appendix~\ref{append} and setting the background to AdS space, using \equ{ADSRiemann}, we can show that $\Q_{zzij}= \frac{1}{4} \K_{ik}\K^{k}_{j}$ and as a result the $\frac{\epsilon}{z}$ divergence exactly vanishes. However the $\frac{\epsilon^2 }{z^2}$ divergence will remain. 

The final step is to take the $\epsilon, z \rightarrow 0$ limit. Depending on the ordering one chooses, there are two ways to do this.  One way is to take $z\rightarrow 0$ limit first. Physically, this corresponds to looking for a divergence in the bulk equation of motion while there is still a small but non-zero conical deficit parameter $\epsilon$.  The second way is take $\epsilon \rightarrow 0$ first. The limit is, therefore, an iterated limit -- the final result depends on the order in which the limit is taken, so there is an inherent ambiguity in this procedure. In fact, this ambiguity can be made even larger in scope if we take the limit simultaneously in $\epsilon$ and $z$. Mathematically, the divergence is a function of the two variables: $\epsilon$ and $z$. In this $\epsilon$-$z$ plane there are an infinite number of paths along which we can take the limit. At least on a mathematical level, there exists no reason why the limit should only be taken along the $z = 0$ or $\epsilon = 0$ path.

The path $z = 0$ is, however, the simplest way to take the limit so as to obtain $\frac{\epsilon}{z} \rightarrow \infty$. In this case, all terms containing $\frac{\epsilon}{z}$ are leading divergences while terms containing $e^{-2\rho(z,\bar z)} \epsilon/z= (z\zb)^\epsilon \epsilon/z$ contribute to sub-leading divergences. Therefore, in this way of taking limits, setting the $H_{zz}$ divergence to zero will yield two different conditions for the minimal surface. The first condition which, after adding the Einstein term, corresponds to the surface equation of motion is
\be
	\K + L^2\lambda(\R\K-2 \K^{ij} \R_{ij})= 0\,.
\label{GGEeom}
\ee
This agrees with the surface equation that comes from the Jacobson-Myers functional. However, there will also be an extra constraint~\cite{erd} of the form  
\be
	-\K^3+3\K\K_{2}-2\K_3=0\,,
	\label{ExC}
\ee
coming from the sub-leading divergence. The $H_{ij}$ divergence will also lead to a similar constraint. The above condition can only be true for very special surfaces and therefore is an over-constraint on the surface. In fact, if these two conditions were to be true simultaneously, the surface equation of motion we would end up getting is:
\be
  c\,\K  + \alpha \lambda L^2 (\K^3-3\K\K_{2}+2\K_3)\, =0.
\ee
To get this form of the equation, we have used the Gauss identities on AdS space. Here, $c = (1-2f_{\infty}\lambda)$ is proportional to the Weyl anomaly and $\alpha$ is a variable that can take any arbitrary numerical value. The surface equation of motion corresponding to the Jacobson-Myers functional can be recovered if $\alpha=1$. However, at present nothing within the LM method sets the value of this parameter to one. Note that if $\alpha$ was zero, the minimal surface that we would get is the same as in the Einstein case. It also the minimal surface that would follow if one were minimizing just the Wald part of the entropy functional. 

In the above paragraph we outlined one way in which the LM method could potentially give rise to the correct surface equation of motion. Let us now explore if we can change the limit-taking procedure itself to get the correct equation. This can be accomplished by choosing a different path in the $\epsilon$-$z$ plane to take the limit. Taking the limit along the path $\epsilon = 0$ will simply kill off all divergences; this is not surprising since physically this corresponds to turning off the conical deficit in the metric. 
However, we can pick a path in the $\epsilon, z$ plane that will kill off the sub-leading divergence but preserve the leading divergence. For example, as was shown in Ref.~\cite{aks}, taking any path of the form $(z)^{2\epsilon}={(\frac{z}{\epsilon})}^{1+v}$, where $v$ is a number greater than one, will keep only the leading divergence. At this point, we can offer no justification of why one should choose this particular way of taking limits. We are merely demonstrating that there does exist a way to take limits in the $\epsilon,z$ plane that leads to the correct surface equation of motion in the Gauss-Bonnet case. This way of taking limits is equivalent to discarding terms suppressed by $e^{-2\rho(z,\bar z)}$ and was also used in Ref.~\cite{dong} to show that the LM method leads to the same surface equation of motion in the Lovelock case as can be derived from the entropy functional for Lovelock gravity~\cite{dong,Kang,Sarkar}. This is the way of taking limits that we will use. However, unless one can specify a mechanism or a physical interpretation which reproduces this way of taking limits (which is possible if the metric itself is re-defined), the argument that the LM method reproduces the correct surface equation of motion for Gauss-Bonnet theory remains incomplete.

The same ambiguity in taking limits exists for general R$^2$ theory. To remain consistent with the Gauss-Bonnet point, for R$^2$ theory we will continue to take limits as stated in the paragraph above. However, in the general case this is not an ideal solution. As we will see, $\sim \K^3$ terms always occur with the $e^{-2\rho(z,\bar z)}$ factor in the divergence equations for R$^2$ theory. This means that if we use the above way of taking limits we will never get such terms at any point in the parameter space. As we saw in \equ{R2minimal}, the surface equation of motion for R$^2$ theory does contain such terms. However, our goal for general R$^2$ theory is to see to what extent we are able to reproduce the surface equation of motion in \equ{R2minimal}, while taking the limit in such a manner that the result at the Gauss-Bonnet point agrees with what comes from the Jacobson-Myers functional. 
It is clear, though, that the question of taking limits in the LM method deserves more study.

\subsubsection{The general case \label{newmetric}}
We now work out the divergence equations for the R$^2$ case.
For general R$^2$ theory all second-order quantities will enter into the divergences. We can anticipate the effect that terms containing $\Omega$ and $\B$ will have on the surface equation of motion coming from the LM method. Consider the following components of the bulk Riemann tensor around $z=\zb=0$:
\begin{align}
R_{pqrs}\Big|_{(z=0,\zb=0)} &= -3 e^{4\rho(z,\bar z)} \hat\ve_{pq} \hat\ve_{rs} \O \,,\nn\\
R_{pqri}\Big|_{(z=0,\zb=0)} &= 3 e^{2\rho(z,\bar z)} \hat\ve_{pq} \B_{ri} \,,
\label{Riemanns1}
\end{align}
where, $\hat\ve_{ab}$ is defined as $\hat\ve_{z\zb}=-\hat\ve_{\zb z}= e^{-2\rho(z,\bar z)} g_{z\zb}$. The quantity $\Omega$ is therefore equivalent to $-1/3 R_{\mu\nu\rho\sigma}n^{\mu}_{r}n^{\nu}_{s}n^{\rho}_{r}n^{\sigma}_{s}$ evaluated at $z,\zb=0$. We can determine a numerical value for the quantities  $\B^{r}_{i} $ and $\O$ in the metric by demanding that the bulk Riemann tensor be the AdS solution at zeroeth order. Since for AdS space the Riemann tensor is given by \equ{ADSRiemann}, we can write the components of the bulk Riemann tensor on the L.H.S of \equ{Riemanns1} in terms of the components of the bulk metric. Expanding the metric using \equ{metric2} and keeping only the zeroeth-order terms in $z,\zb$ we get
\be
	\O = -\frac{1}{12} C \;\;\; \textrm{and}  \;\;\; \B_{ri} = 0
	\label{Tvalue}
\ee
Therefore, $\B^{r}_{i}$ can be set to zero. In writing the divergences, we also ignore\footnote{As we saw for the Gauss-Bonnet case, these terms will be present in the divergences, but they will not change our conclusions for $R^2$ theory.} $Q^{zz}_{ij}$ and $Q^{\zb\zb}_{ij}$, the remaining component being $Q_{ij}=Q^{\zb z}_{ij}=Q^{z\zb}_{ij}$. 

For R$^2$ theory the derivative order of the equation of motion is four. That means we should include order $z^3$ terms in the metric, since they can contribute to the divergences. These terms can be parametrized as
\be
ds^2 =  e^{4\rho(z,\bar z)} \Delta_{pqrst} x^p x^q x^r dx^s dx^t + \W_{rspij} x^r x^s x^p  dy^i dy^j + 2 e^{2\rho(z,\bar z)}  \C_{rsi} x^r x^s (\zb dz-z d\zb) dy^i \,.
\label{metric3}
\ee
This is the most general form of the third-order terms in the metric. Here, we have written the $e^{2\rho(z,\bar z)} $ dependence of each term explicitly. As for the second-order quantities, the third-order quantities $ \Delta_{pqrst}, \W_{rspij} $ and $\C_{rs i} $ can be found by calculating the curvature tensors, but to linear order in $z(\zb)$. Then, for example, $e^{4\rho(z,\bar z)}\Delta_{pqrst} \equiv -1/6 \partial_p ({R_{\mu\nu\rho\sigma}n^{\mu}_{q}n^{\nu}_{r}n^{\rho}_{s}n^{\sigma}_{t}})$ evaluated at $z=\zb=0$. Note that the factor of $e^{4\rho(z,\bar z)}$ will cancel from both sides on using the AdS background. In fact, this particular term vanishes altogether in this background. On using the metric with the third-order terms listed above to find the divergences in the equation of motion we find that the $\C_{rs i} $,  $\W_{zzzij}$ and $\W_{\zb\zb\zb ij}$ do not contribute. The terms that are relevant are $\W_{zz\zb ij}$ and $\W_{\zb\zb z ij}$, because as will show below they will lead to unsuppressed $C \K$ terms. Without loss of generality, we can set them to be equal and denote this term as $\W_{ij}$. 

For general R$^2$ theory, the divergence in the $zz$ component from the $H_{\a\b}$ term in the bulk eom is 
\begin{align}
~~~~~H_{zz} ~=~\frac{\epsilon }{ z}&\Big[-\tfrac{1}{2}(\lambda_{2}+4\lambda_{3})\nabla ^2\K+(2\lambda_1+\lambda_2+2\lambda_3)\nabla^i\nabla^j\K_{ij}+\lambda_{3}(\R\K-2 \K^{ij} \R_{ij})~+~\nonumber\\&~~~ 4(-2\lambda_1+3\lambda_2+14\lambda_3)\K_{ij}\mathcal{A}^i\mathcal{A}^j-6(\lambda_2+4\lambda_3)\K \mathcal{A}^i\mathcal{A}_i~+\nonumber\\&~~~
  8(3\lambda_1+2\lambda_2+5\lambda_3)\K \Omega  \Big]~-\nonumber\\\frac{\epsilon}{z^2}&\Big[e^{-2\rho(z,\bar{z})}\Big\{(2\lambda_{1}-\lambda_{2} -6\lambda_{3} ) \K_2+ \tfrac{1}{2}(\lambda_{2}+4\lambda_{3})\K^2+2(\lambda_2+4\lambda_3) \Q\Big\}\Big]~+\nonumber\\
\frac{\epsilon}{z}&\Big[e^{-2\rho(z,\bar{z})}\Big\{-\lambda_{3}\K^3+(\lambda_{2}+7 \lambda_{3})\K\K_{2}-2(3\lambda_{1}+2\lambda_{2}+6\lambda_{3})\K_3~+\nonumber\\
&~~~~~~~~~~~~~~~~(6\lambda_{1}+5\lambda_{2}+14\lambda_{3})\K_{ij}\Q^{ij}-\tfrac{3}{2}(\lambda_{2}+4\lambda_{3})\K\Q~-\nonumber\\&~~~~~~~~~~~~~~~~~4(\l_2 +4 \l_3) \W\Big\}\Big] \,.
\label{zzcomponentm2}
\end{align}
The divergences in the other components are
\begin{align}
\hspace{-.5cm}H_{z\bar z}~= ~\frac{2 \epsilon}{z}&\Big[ e^{-2\rho(z,\bar z)}\Big\{(\lambda_{3}+\tfrac{1}{4}\lambda_2)\K^3+(\lambda_1-\tfrac{3}{2}\lambda_2-7\lambda_3)\K\K_{2}\nonumber~+\\&~~~~~~~~~~~~~2(-2\lambda_1+\lambda_2+6\lambda_3)\K_{3}+(2\lambda_1-3\lambda_2-14\lambda_3)\K_{kl}\Q^{kl}\nonumber~+\\&~~~~~~~~~~~~~\tfrac{3}{2}(\lambda_2+4\lambda_3)\K\Q +  8( \lambda_2+4\lambda_3) \W \Big\}\Big]\,,
\label{zbzcomponentm2}
\end{align}
\begin{align}
~~~~~H_{zi}~=~ \frac{2\epsilon }{z}&\Big[e^{-2\rho(z,\bar z)}\Big\{-\tfrac{1}{2}(2\lambda_1+\lambda_2)\K^{k}_{i}\nabla_{k}\K-(3\lambda_1-\lambda_{2}-6\lambda_3)\K^{kl}\nabla_{i}\K_{kl}~-\nonumber\\&~~~~~~~~~~~~~\tfrac{1}{4}(3\lambda_2+8\lambda_3)\K\nabla_i\K+
(5\lambda_{1}+\lambda_{2})\K^{k}_{i}\nabla_{l}\K^{l}_{k}-\lambda_{1}\K\nabla_{k}\K^{k}_{i}
~+\nonumber\\&~~~~~~~~~~~~~~(9\lambda_1+2\lambda_2)\K^{kj}\nabla_{k}\K_{ji}-(\lambda_{2}+4\lambda_{3})\nabla_{i}\Q-(4\lambda_{1}+\lambda_{2})\nabla_{k}\Q^{k}_{i}
~-~\nonumber\\&~~~~~~~~~~~~~~(10\lambda_1-2\lambda_2-18\lambda_3) \mathcal{A}_i\K_2-\tfrac{1}{2}(3\lambda_2+12\lambda_3) \mathcal{A}_i\K^2~+~\nonumber\\&~~~~~~~~~~~~~~8(4\lambda_1+\lambda_2)\K_{ij}\K^{jk}\mathcal{A}_{k}
-2(\lambda_2+4\lambda_3)\mathcal{A}_{i}\Q\Big\}\Big]\,,
\label{zicomponent}
\end{align}
\begin{align}
~~~H_{ij}~=~\frac{4\epsilon }{z}&\Big[e^{-4\rho(z,\bar z)}\Big\{(\tfrac{1}{3}\lambda_{1}+\tfrac{1}{4}\lambda_{2}+\tfrac{2}{3}\lambda_3)h_{ij}\K^3-(7\lambda_1+2\lambda_2+2\lambda_3)\K\K_{ik}\K^{k}_{j}~+\nonumber\\&~~~~~~~~~~~~~2(16\lambda_{1}+4\lambda_{2}+\lambda_3)\K_{ik}\K^{kl}\K_{lj}-(\lambda_1+3\lambda_2+10\lambda_3)h_{ij}\K\K_2~-\nonumber\\&~~~~~~~~~~~~~
(3\lambda_1-2\lambda_3)\K_{ij}\K_2-\tfrac{1}{3}(\lambda_1-18\lambda_2-70\lambda_3)h_{ij}\K_3~+\nonumber\\&~~~~~~~~~~~~~2(4\lambda_{1}+\lambda_{2})\Q_{ij}\K+2(\lambda_{2}+4\lambda_{3})h_{ij}\K\Q-8(4\lambda_{1}+\lambda_2)\K_{ik}\Q^{k}_{j}
~-\nonumber\\&~~~~~~~~~~~~~(\lambda_{2}+4\lambda_{3})\K_{ij}\Q-7(\lambda_{2}+4\lambda_{3})h_{ij}\K_{kl}\Q^{kl} + 32 (4\lambda_1+\lambda_2) \W_{ij} + \nn \\
&~~~~~~~~~~~~~32(\lambda_2+4\lambda_3)h_{ij} \W \Big\}\Big]\,.
\label{ijcomponent}
\end{align}
Whether or not the divergences in the $ij$, $zi$ and $z\zb$ components vanish before taking the $\epsilon\rightarrow 0$ limit will depend upon the exact values of the second-order terms. The $zi$ divergence, in particular, should be equivalent to the constraint equation coming from the tangential variations and should vanish by the Codazzi relation in \equ{codazzi}. As in the Gauss-Bonnet case, the divergence in the $ij$ component is not expected to fully vanish by itself. We therefore take the limit as prescribed in the last section. This reduces the divergences in the $zi$ and $ij$ components to zero. However, because of the presence of the $\W$ term there still remains an unsuppressed divergence in the $z\zb$ component. This divergence can only go to zero if $\K=0$ or the theory is at the Gauss-Bonnet point.

We now examine the divergences in the $zz$ component, to be able to compare it with the surface equation of motion derived using the FPS functional. First looking at the $1/z$ divergence in that component, one can see that it contains the unsuppressed terms $\K_{ij} \A^i \A^j$ and $\K \A^i \A_i$ which are not present in \equ{R2minimal}. However, these terms can be eliminated in favor of other variables. Consider the $R_{zi\zb j}$ component of the Riemann tensor expanded around $z=0,\zb=0$:
\be
  R_{zi\zb j}\Big|_{(z=0,\zb=0)} = \tfrac{1}{2} e^{2\rho(z,\bar z)} \mathcal{F}_{ij} - 2 e^{2\rho(z,\bar z)} \A_i \A_j  + \tfrac{1}{4} \K_{zik} \K^k_{\zb j} -\tfrac{1}{2} \Q_{z \zb ij} \,.\label{Riemanns21}
\ee
Using \equ{ADSRiemann} again and multiplying both sides by $\K_{ij}$, we find that the $\mathcal{A}_i \mathcal{A}_j \K^{ij}$ term can be written as $\sim C\K + e^{-2\rho(z,\bar z)} \K^3 + e^{-2\rho(z,\bar z)} \Q\K$. The $\A_i \A^i \K$  terms can be written in a similar fashion. Since only the $C\K$ term is unsuppressed we find
\begin{align}
	\mathcal{A}_i \mathcal{A}_j \K^{ij} &= \frac{C\K}{4} + \cdots\;\; {\rm and } \nn \\
	\A_i \A^i \K &= \frac{3C\K}{4} + \cdots\,,
	\label{a2kvalue}
\end{align}
where the dots denote the suppressed terms. 

Next looking at the $e^{-2\rho(z,\bar z) }/z$ divergence we find that the $\W$ term will contribute to the surface equation of motion, since this term contains a $e^{2\rho(z,\bar z)}$ factor that enhances the divergence to $1/z$.
This term can be determined by using the following equation
\be
	\partial_z R_{\zb \zb}\Big|_{(z=0,\zb=0)} = - \W + 2  e^{2\rho(z,\bar z)} \K^{ij}\A_i \A_j  - 2  e^{2\rho(z,\bar z)}\Omega \K + \cdots.
\ee
The R.H.S of this equation disappears in the AdS background. Using Eqs.~(\ref{Tvalue}) and (\ref{a2kvalue}) we find
\be
	\W= \frac{2  e^{2\rho(z,\bar z)} C\K}{3} + \cdots\,.
\ee
The $\Q$ terms that are also present in this divergence do not contribute since as we show below they are expected to contain only $\sim \K^2$ terms and therefore remain suppressed.

Substituting these values in \equ{zzcomponentm2}, and adding the Einstein term we find that the $1/z$ divergence of the $zz$ component gives rise to the following surface equation of motion:
\be
\K + L^2\{(2\lambda_{1}+\tfrac{1}{2}\lambda_{2})\nabla ^2\K+\lambda_{3}(\R\K-2 \K^{ij} \R_{ij}) + \l_1 C_1 \K + \l_2 C_2 \K + \l_3 C_3\K\}=0\,
\label{GGEeom2}
\ee
where $C_1=-4C$, $C_2=-11 C/2 $ and $C_3=-18 C $. The coefficients of the $\nabla^2\K, \R\K$, $\K^{ij} \R_{ij}$ and $C\K$ terms in the above equation all match with those in \equ{R2minimal}. Because of the way we are taking limits, the $\K^3$ terms that are present in \equ{R2minimal} are not present here.

Finally we look at the $\epsilon/z^2$ divergence in the $zz$ component. For this divergence to vanish, we get the condition
\be
  (2\lambda_{1}-\lambda_{2} -6\lambda_{3} ) \K_2+ \tfrac{1}{2}(\lambda_{2}+4\lambda_{3})\K^2+2(\lambda_2+4\lambda_3) \Q =0 \,.
  \label{ExtraDiv}
\ee
To satisfy this condition at arbitrary points of the parameter space, one has to demand that $\Q$ be a function of $\sim \K^2$ terms, and also $\l_1$, $\l_2$ and $\l_3$. Demanding that $\Q$ be independent of  $\l_1$, $\l_2$ and $\l_3$, will pick out a special point in the parameter space (apart from the Gauss-Bonnet point where this condition is trivially satisfied). 

To summarize the results for R$^2$ theory:
\begin{enumerate}
 \item Apart from the absence of $\sim\K^3$ terms, \equ{GGEeom2} that we found using the LM method is exactly the surface equation of motion that results from the FPS functional.
 \item There are some problematic extra divergences. The $z \zb$ component of the bulk equation of motion has a divergence that can only disappear at the Gauss-Bonnet point. There is also a second-order $1/z^2$ divergence in the $zz$ component. This can be taken to fix the value of the term $\Q$; however, it is not possible to do this in a way that is independent of the parameters of R$^2$ theory.
\end{enumerate}

\subsection{The stress-energy tensor from the brane interpretation \label{deltas}}
In Ref.~\cite{maldacena}, it was noted that a equation of motion of a cosmic string is the same as the equation for the minimal entangling surface. This is because a cosmic string produces a spacetime with a conical defect with a metric of the form in \equ{metric2}. The equation of motion is given by minimizing its action. For Einstein gravity this is just the Nambu-Goto action and equation of motion of a cosmic string is
\be
    \K = 0.
\ee
This condition minimizes the surface area of the string as it sweeps through spacetime. The same thing holds for a cosmic brane.

As was done in Ref.~\cite{dong}, where it was referred to as the cosmic brane method, this fact can be exploited to construct the entropy functional from the bulk equation of motion. In this section, we will check this construction of Ref.~\cite{dong}. The idea is that the bulk equation of motion in \equ{bulkeom} should lead to the cosmic brane as a solution, to linear order in $\epsilon$. In particular, this means that L.H.S of \equ{bulkeom} should be equal to the stress-energy tensor of the brane. Since the brane is a localized source, the stress-energy tensor will contain delta functions. Once we have found the stress-energy tensor we can identify the associated action via $T_{\alpha \beta} = \frac{ \delta{S}}{\delta g_{\alpha \beta}}$. 

Let us see how this works in the Gauss-Bonnet case. In the bulk equation of motion, terms such as $\partial_{\bar{z}}\partial_{z}\rho(z,\bar z)$ correspond to delta functions. We set $\delta(z,\bar z) = e^{-2\rho(z,\bar{z})} \partial_{\bar{z}}\partial_{z}\rho(z,\bar z)$. Note that $\delta(z,\bar z)$ defined this way contains a factor of $\epsilon$. 

The delta divergences in the $ij$ component of the bulk equation of motion to linear order in $\epsilon$ are then:
\begin{align}
T_{ij} =\delta(z,\bar z)\Big\{  &-4 \, \lambda \,(h_{ij} \R -2 \R_{ij}) +\nn \\
& -2 \,\lambda\, e^{-2\rho(z,\bar{z})}(h_{ij}\K_{2}- h_{ij}\K^2+2\K_{ij}\K-2\K_{ik}\K^{k}_{j} )\Big\}\,.
\end{align}
To identify this as the stress-energy tensor coming from the Jacobson-Myers functional (interpreted as a cosmic brane action), the second term should go to zero. This term carries a factor of $e^{-2\rho(z,\bar{z})}$ as compared to the first term and according to our way of taking limits is suppressed.  
Our result is then in agreement with the claim in Ref.~\cite{dong} that the cosmic-brane method can be used to show that the Jacobson-Myers functional is the right entropy functional for Gauss-Bonnet theory. 

However, as we will see there are problems for the general four-derivative theory. For R$^2$ theory, the delta divergences in the $ij$ component are
\begin{align}
~T_{ij}~=~ &\delta(z,\bar z) \Big[ -4 \,\lambda_3 \,\,(h_{ij} \R -2 \R_{ij}) - 16 (6\lambda_1+11\lambda_2+38\lambda_3)h_{ij} \Omega ~+
\nonumber\\&~~~~~~~ ~~~  e^{-2\rho(z,\bar z)}\Big\{-(\lambda_2+2\lambda_3) h_{ij}\K^2+2(\lambda_2+3\lambda_3)h_{ij}\K_2~-
\nonumber\\&~~~~~~~~~~~~~~~~~~~~~~~~~~2(12\lambda_1+4\lambda_2+4\lambda_3)\Q_{ij}-2 (\lambda_2+4\lambda_3)\Q h_{ij}~-
\nonumber\\&~~~~~~~~~~~~~~~~~~~~~~~~~~2(4\lambda_1+\lambda_2+2\lambda_3)\K_{ij}\K+2(14\lambda_1+4\lambda_{2}+4\lambda_3)\K_{kj}\K^{k}_{j} )~+
\nonumber\\&~~~~~~~~~~~~~~~~~~~~~~~~~~16(20\lambda_1+11\lambda_2+24\lambda_3)\mathcal{ A}_{i}\mathcal{A}_{j}\Big\}~ \Big]
+\nonumber\\&e^{-2\rho(z,\bar z)}\Big\{\partial_{z}\delta(z,\bar z)+\partial_{\bar z}\delta(z,\bar z)\Big\}\Big\{-2(2\lambda_{1}+\lambda_2+2\lambda_{3})\K_{ij}+(4\lambda_3+\lambda_2) h_{ij}\K\Big\}~
-\nonumber\\&4e^{-2\rho(z,\bar z)}\,\partial_{z}\partial_{\bar z}\delta(z,\bar z)(\lambda_2+4\lambda_3)\,.
\end{align}
Again, barring the term suppressed by $e^{-2\rho(z,\bar{z})}$, we have checked that the result for this component is of the same form as that produced on calculating the stress-energy tensor from an action equivalent to the FPS functional. The derivative of delta terms like $\partial_{z}\delta(z,\bar z)$ are typical in the stress-energy tensor of actions containing terms that depend on the extrinsic curvature \cite{Stringcurv2}.  
However, the $zz$  and $z\zb$ components of the bulk equation of motion also contain delta divergences that are not suppressed:
\be
T_{zz}~=~-4\partial_{z}^{2}\delta(z,\bar z)(2\lambda_{1}+\lambda_{2}+2\lambda_{3})-2 \partial_{z}\delta(z,\bar z)(4\lambda_{1}+\lambda_2)\K
\ee
and
\begin{align}
T_{z\bar z}~=~-2 &\big\{\partial_{z}\delta(z,\bar z)+\partial_{\bar z}\delta(z,\bar z)\big\}\big(2\lambda_1+\lambda_2+2\lambda_3\big)\K~+\nonumber\\&4\,\partial_{z}\partial_{\bar z} \delta (z,\bar z)(2\lambda_1+\lambda_2+2\lambda_3)\,.
\end{align}
Taking the delta divergences in all components into account, the $T_{\mu \nu}$ we have found does not look like the stress-energy tensor for a cosmic brane corresponding to a three-dimensional surface in the five-dimensional bulk. Note that the extra divergences all vanish for the Gauss-Bonnet theory. The Gauss-Bonnet result therefore stands. However, any attempt to use this method to show that the FPS functional is the correct entropy functional for R$^2$ theory should be able to account for these extra delta divergences.   

\section{Quasi-topological gravity \label{Quasi}}
The lagrangian for quasi-topological gravity~\cite{quasi} contains terms cubic in the Riemann tensor. It can be used to study a class of CFT's involving three parameters in four dimensions. It has many interesting features including the fact that its linearized equation of motion is two-derivative order. Unitarity for this theory was studied in Ref.~\cite{Sisman}.

In Sec.~(\ref{qef}), we find the HEE functional for quasi-topological theory using \equ{eei} and compute the universal terms is Sec.~(\ref{uef}). In Sec.~(\ref{eomef}), we find the surface equation of motion for this theory using the LM method.
\subsection{The entropy functional \label{qef}}
The action for quasi-topological theory in five dimensions is 
\be
S_{QT} =-\frac{1}{2\lp^3}\int d^5 x \Big( \mathcal{L}_1 + \mathcal{L}_2 + \nu \,Z_{5}\Big)\,,
\ee
where $\mathcal{L}_1$ is the Einstein-Hilbert action given in \equ{Ei} and $\mathcal{L}_2$ is the Gauss-Bonnet lagrangian as in \equ{Lr} with $\lambda_1=\lambda_3=\lambda\,, \lambda_2=-4\lambda$.
The last term is the R$^3$ lagrangian:
\begin{align}
Z_5=&\mu_{0} R_{\a\b}{}^{\gamma\delta}R_{\gamma\delta}{}^{\mu\nu}R_{\mu\nu}{}^{\a\b}+\mu_1 R_{\a}{}^{\b}{}_{\gamma}{}^\delta 
R_{\b}{}^{\eta}{}_{\delta}{}^{\zeta} R^{\a}{}_{\eta}{}^{\gamma}{}_{\zeta} +
\mu_2 R_{\a \b \gamma \delta}R^{\a \b \gamma \delta} R~+~\nn \\&\mu_3 R_{\a \b \gamma 
\delta}R^{\a
\b \gamma}{}_{\eta}R^{\delta \eta}+\mu_4 R_{\a \b \gamma \delta} R^{\a 
\gamma}R^{\b \delta}
+\mu_5 R_\a{}^{\b}R_\b{}^{\gamma}R_\gamma{}^{\a} +\mu_6 
R_\a^{\,\,\b}R_\b^{\,\,\a}R+\mu_7 R^3 \label{Lrr} \,.
\end{align}
There are two different consistent R$^3$ theories. For the first theory
\be \label{th}
\mu_0=0\,,~
\mu_{1}=1\,,~\mu_{2}=\tfrac{3}{8}\,,~\mu_{3}=-\tfrac{9}{7}\,,~\mu_{4}=\tfrac{15}{7}\,,~
\mu_{5}=\tfrac{18}{7}\,,~\mu_{6}=-\tfrac{33}{14}\,,~\mu_{7}=\tfrac{15}{56}
\ee
and the coupling constant is $\nu=\frac{ 7 \mu L^4}{4}$, while for the second theory
\be \label{th1}
\mu_0=1\,,~
\mu_{1}=0\,,~\mu_{2}=\tfrac{3}{2}\,,~\mu_{3}=-\tfrac{60}{7}\,,~\mu_{4}=\tfrac{72}{7}\,,~
\mu_{5}=\tfrac{64}{7}\,,~\mu_{6}=-\tfrac{54}{14}\,,~\mu_{7}=\tfrac{11}{14}
\ee
and the coupling constant $\nu=\frac{ 7 \mu L^4}{8}\,.$

The R$^3$ part for the HEE functional is
\begin{align} 
\begin{split}
\label{R3}
S_{\rm EE,\, R^3}&= \frac{2\pi \nu}{\lp^{3}}\int d^{3}x \sqrt{h}\, \left(\mathcal{L}_{\rm Wald,\,R^3}+\mathcal{L}_{\rm Anomaly,\, R^3}\right)\,,
\end{split}
\end{align}
where
\beq
\mathcal{L}_{\rm Wald, R^3}&=&6\mu_0 R^{ z \bar z \alpha \beta}R_{  z \bar z \alpha\beta}+3 \mu_1 \big(R^{z \alpha \bar z}{}_{ \beta}R_{z \alpha \bar z }{}^{\beta }-R^{z \alpha z}{}_{ \beta}R_{ z \alpha z}{}^\beta\big)
+\mu_2\big(R_{\alpha\beta\rho\sigma}R^{\alpha\beta\rho\sigma}-\nonumber \\
&&4 R\,R^{z \bar z }{}_{\bar z  z}\big)+2\mu_3 \big(R_{\alpha}{}^{z}{}_{\bar z }{}^{\bar z} R^{ \alpha}{}_{  z}-R_{\alpha}{}^{\bar z}{}_{z}{}^{z} R^{ \alpha}{}_{\bar z}+ \tfrac{1}{2} R_{\alpha\beta\rho }{}^{\bar z} R^{\alpha\beta\rho }{}_{\bar z}\big) +
\mu_4 (2 R^{z}{}_{\alpha}{}_{ z}{}_{ \beta}R^{\alpha\beta}+\nonumber\\
&& 
(R^{z}{}_{ z})^2- R^{zz}R_{ z  z})+  3 \mu_5 R^{z \alpha}R_{z\alpha}+
\mu_6 \big (R_{\alpha\beta}R^{\alpha\beta}+2 R R^{z }{}_{z}\big)+
3 \mu_7 R^2\,.
\eeq
The symbols $z$ and $\bar z$ in the above expression label the two orthogonal directions while the indices $\alpha ,\beta,...$ are the usual bulk indices. The expression for the anomaly part is
\begin{align}
\mathcal{L}_{\rm Anomaly,\,R^3}=&\mu_0(12\, \K_2{}^{ij} \Q_{ij}-6\, \K_4)-\mu _1(\tfrac{3}{2}\,\K_4-\tfrac{3}{2}\K_2{}^2+3\,\K_{ij}\K_{kl}\R^{ikjl})-\nonumber\\&\mu _2 (6\, \K_2{}^2-2\, 
\K_2\,\K^2-8\, \K_2\,\Q+4\, \K_2 \R)-\nonumber\\&\mu _3(2\, \K_4+\tfrac{1}{2}\, 
\K_2{}^2-\K_2\,\Q-2\, \K_2{}^{ij} \Q_{ij}-2\, \K^{ij}\Q_{ij}\,\K+2
\,\K_2{}^{ij}\R_{ij})-\nonumber\\&\mu _4(2\, \K_3 \,\K-\K_2\,\K^2-2 
\K^{ij}\Q_{ij}\,\K+2\, \K^{ij}\R_{ij}\,\K)-\nonumber\\&\mu _5 (\tfrac{3}{4}\, \K_2 
\,\K^2-\tfrac{3}{2}\, \K^2 \,\Q)-\mu _6 (\tfrac{3}{2} \,\K_2\,\K^2-\tfrac{1}{2} 
\K^4-2\, \K^2\Q+ \K^2\R)\,.
\end{align}
where $\K_{4}=\K^{ij}\K_{jl}\K^{lk}\K_{ki}\,.$
In calculating the anomaly part from \equ{eei}, we have used the value of $\B^i=0$ that we found in Sec.~(\ref{newmetric}). This is the reason that while terms involving $\B^i$ are supposed to contribute to \equ{eei}, the above equation does not contain any terms containing $\B^i$. The full HEE functional has contributions from the Einstein and R$^2$ part also which are given in~Eqs.(\ref{Wald}) and (\ref{Extra}).
\subsection{Universal terms \label{uef}}
In this section, we will demonstrate that our HEE functional for the quasi-topological gravity produces the correct universal terms. For the general structure and calculation of the universal term of the entanglement entropy in four dimensions, see \cite{solod1}. These central charges can be easily calculated using the technique of Ref~\cite{anom}. 

We follow the procedure given in \cite{abs} for R$^2$ theory. Here we sketch the main steps of this calculation. We will minimize \equ{R3} for a bulk surface with a spherical and cylindrical boundary. We will carry out this procedure for the five-dimensional bulk AdS metric 
\be \label{metric}
ds^{2}=\frac{\tilde L^{2}}{ z^{2}}(dz^{2}+d\tau^{2}+h_{ij}dx^{i}dx^{j})\,.
\ee
Here, $\tilde L$ is  the AdS radius and $h_{ij}$ is a 3-dimensional boundary metric 
given below.  
For the calculation of EE for a spherical entangling surface we can write the boundary $h_{ij}$ in 
spherical polar coordinates as
\be
^{sphere}h_{ij}dx^{i}dx^{j}= d\rho^{2}+\rho^{2}d\Omega_{2}^{2}\,,
\ee 
where $d\Omega_{2}^2=d\theta^{2}+\sin^2 \theta d\phi^{2}$ is the metric of a 
unit two-sphere, $\theta$ goes from $0$ to $\pi$ and $\phi$ goes from $0$ to $2\pi$.
Similarly, for a cylindrical entangling surface
\be
^{cylinder}h_{ij}dx^{i}dx^{j}=du^{2}+d\rho^{2}+\rho^{2}d\phi^{2}\,.
\ee
Here, $u$ is the coordinate along the direction of the length of the cylinder. For a 
cylinder of length $H$, $u$ goes form $0$ to $H\,.$

We set $\rho=f(z), \tau=0$ in the metric in \equ{metric}  and 
minimize the entanglement entropy functional (whose R$^3$ part is given in \equ{R3}) on this codimension 2 surface to find the Euler-Lagrange 
equation for $f(z)$. Using the solution for $f(z)$ we evaluate the entropy functional to 
get the EE.
  
For the spherical boundary, we get  $f(z)=\sqrt{f_{0}^{2}-z^{2}}$ which gives the EE as
\be
S_{EE}= -4 a \ln(\frac{f_{0}}{\delta})\,.
\ee
Here, $\delta$ is the UV cut-off that comes from the lower limit of the $z$ integral and 
$f_{0}$ is the radius of the entangling surface. The value of $a$ is 
 \begin{align} 
 \label{a} a&= \frac{\pi^{2} 
L^{3}}{f_{\infty}^{3/2}\lp^{3}}(1-6f_{\infty}\lambda+9 f_{\infty}^2\mu)\,. 
\end{align}
For this case, the entire contribution comes from the Wald entropy as the extrinsic curvatures are identically zero.

For the cylindrical boundary, we find $f(z)=f_{0}-\frac{z^{2}}{4f_{0}}+...$ leading to
\be
S_{EE}= - \frac{c H}{2 R} \ln(\frac{f_{0}}{\delta})\,. 
\ee
The  value of $c$ corresponding to the theory in \equ{th} is
 \begin{align} 
 \label{c1}
c&= \frac{\pi^{2} 
L^{3}}{f_{\infty}^{3/2}\lp^{3}}\Big\{1-2f_{\infty}\lambda+9 f_{\infty}^2\mu+ f_{\infty}^2\mu (42\mu_{1} - 336 \mu_2 - 56 \mu_3)\Big\}\,,
\end{align}
while that corresponding to the theory in \equ{th1} is
\begin{align} 
\label{c2}
c&= \frac{\pi^{2} 
L^{3}}{f_{\infty}^{3/2}\lp^{3}}\Big\{1-2f_{\infty}\lambda+9 f_{\infty}^2\mu- f_{\infty}^2\mu (168 \mu_2 + 28 \mu_3)\Big\}\,.
\end{align}
The $1+9 f_{\infty}^2\mu$ part is the usual Wald entropy contribution, while the remaining part comes from the anomaly part. After putting in the values of $\mu$'s given in Eqs.(\ref{th}) and (\ref{th1}) we obtain
\begin{align} 
\label{c}
c&= \frac{\pi^{2} 
L^{3}}{f_{\infty}^{3/2}\lp^{3}}(1-2f_{\infty}\lambda-3 f_{\infty}^2\mu )
\end{align}
for both theories.

These results for the universal terms agree with those calculated in Ref.~\cite{abs} for the two quasi-topological theories. Note from Eqs.~(\ref{c1}) and (\ref{c2}) that only a few terms from $\mathcal{L}_{\rm Anomaly,\, R^3}$ have contributed to the universal term. Terms of the form $\sim\K^4$ do not contribute to this calculation at all. Since $\Q \sim \K^2$, terms of the form $\sim\K^2 \Q$ also do not contribute. 

\subsection{Minimal surface condition \label{eomef}}

We now find the surface equation of motion for quasi-topological  gravity using the LM method. For ease of calculation, we set all second-order quantities and cross-components in the metric in \equ{metric2} to zero. The bulk equation of motion for this theory is \cite{Sinha:2010pm}:
\be
R_{\a\b}-\frac{1}{2}g_{\a\b}R- \frac{6}{L^2}g_{\a\b}-\frac{L^2}{2}H_{\a\b}-\nu F_{\a\b}=0\,,
\ee
where $F_{\a\b}$ is defined in Ref.~\cite{quasi, Sinha:2010pm}.
The $\frac{\epsilon}{z}$ divergence in the $zz$ component of the equation of motion coming from the $F_{\a\b}$ term is
\begin{align}
F^{1}_{zz}=
&\frac{\epsilon}{z}\Big[(\tfrac{3}{2}\mu_1-\mu_2-\mu_3-\tfrac{3}{2}\mu_5-4\mu_6-12\mu_7)\R^{ij}\nabla^2\K_{ij}-(\tfrac{1}{2}\mu_2+\mu_6+6\mu_7)\R^{ij}\K_{ij}\R~+~\nn\\&~~~~(\mu_2+\tfrac{1}{6}\mu_6+3\mu_7)\R_{ij}\R^{ij}\K
+\tfrac{1}{2}(\mu_6+\tfrac{1}{2}\mu_4)\K\nabla^2\R+(\mu_4+\mu_3+4\mu_2)\nabla_i\nabla^i\K~-~\nn\\&~~~~(3\mu_1-8\mu-2-3\mu_3-\mu_4+\tfrac{3}{2}\mu_5)\nabla^l\R_{lijk}\nabla^k\K^{ij}-\tfrac{1}{2}(\mu_4+3\mu_1)\K^{kl}\nabla^i\nabla^j\R_{kijl}~-~\nn\\&~~~~(\tfrac{3}{4}\mu_1-\tfrac{5}{2}\mu_2-\mu_3-\tfrac{1}{2}\mu_4-\tfrac{3}{4}\mu_5-2\mu_6-6\mu_7)\R\nabla^i\nabla^j\K_{ij}~+~\nn\\&~~~~\tfrac{1}{4}(\mu_4+2\mu_3+8\mu_2)\K^{ij}\nabla_{i}\nabla_{j}\R\Big]\,.
\end{align}
While we haven't computed the surface equation of motion that one gets on minimizing the functional in \equ{R3}, this is not very hard to do using the methods of Sec.~(\ref{variation}) and Mathematica\footnote{We have used the Xact package for a number of calculations in this paper}.  The main point is, however, that the surface equation of motion that one will get from the entropy functional will contain $~\K^4$ terms that are absent in the above divergence.

Other divergences are also present in the $zz$ component:
\begin{align}
F^{2}_{zz}=&\frac{\epsilon}{z^2}\Big[ e^{-2\rho(z,\bar z)}\Big \{\tfrac{1}{2}( 3\mu_1+19\mu_2+2\mu_3+\tfrac{14}{3}\mu_6)\R_{ijkl}\K^{ik}\K^{jl}-(\tfrac{7}{2}\mu_6+18\mu_7) \R\K_2 ~+~\nn\\&~~~~~~~~~~~~~~~~(2\mu_2+\tfrac{3}{2}\mu_6+6\mu_7)\R\K^2 +(\mu_4-\tfrac{3}{2}\mu_5)\K\nabla^2\K~-~\nonumber\\&~~~~~~~~~~~~~~~~(\tfrac{4}{3}\mu_2-\mu_4+\tfrac{3}{2}\mu_5+\tfrac{4}{3}\mu_6)\K^{ij}\nabla_{i}\nabla_{j}\K-(\mu_4-\tfrac{3}{2}\mu_5)\K\nabla^i \nabla^j \K_{ij}~+~\nonumber \\&~~~~~~~~~~~~~~~~(3\mu_1-\tfrac{2}{3}\mu_2 -\mu_3 + \mu_4-\tfrac{3}{2}\mu_5-\tfrac{2}{3}\mu_6)
\nabla_{k}\K^{ij}\nabla^{k}\K_{ij}~-~\nonumber\\&~~~~~~~~~~~~~~~~(3\mu_1+\tfrac{2}{3}\mu_2 +\mu_3 +3 \mu_4-\tfrac{3}{2}\mu_5+\tfrac{2}{3}\mu_6)\nabla^i\K_{ij}\nabla^k\K_{k
}^j ~-~\nonumber \\&~~~~~~~~~~~~~~~~(8\mu_2-\mu_4+\tfrac{3}{2}\mu_5)\R_{ij}\K^{ij}\K+(3\mu_1+8\mu_2-\mu_4-2\mu_6)\R_{ij}\K^{jk}\K_{k}^{i}\Big\}\Big]~+~\nonumber\\& 
\frac{\epsilon}{z^3}\Big[e^{-4\rho(z,\bar z)}\Big\{(3\mu_1-2\mu_2-2\mu_3)\K_3+(\mu_4-3\mu_5-2\mu_6)\K\K_2\Big\}\Big]\,.
\end{align}
As for R$^2$ theory, these divergences can be used to determine second and higher-order terms in the metric. At linear-order in the metric, divergences in all other components of the equation of motion go to zero if we take the limit as mentioned in Sec.~(\ref{LMmethod}).

\section{Discussion \label{discussion}}

In this paper, we found the surface equation of motion for general R$^2$ theory and quasi-topological gravity using the generalized gravitational entropy method of Ref.~\cite{maldacena}. We found that these do not match exactly with what can be derived by extremizing the HEE functional for these theories -- the HEE functional being calculated using the formula proposed in Refs.~\cite{dong,camps}.

Let us summarize our findings regarding R$^2$ theory. First, the leading-order terms on both sides do match. In fact, barring $\sim\K^3$ terms, the surface equation of motion that follows from the LM method is precisely the surface equation of motion that follows from the FPS functional. 

The main problem with the LM method is that there are divergences in components other than the $zz$ component, for a general higher-derivative theory. In the Gauss-Bonnet case, there are ways we can take the limit to set these divergences to zero. However, the effect of taking the limit in this way is to remove all $\sim\K^3$ divergences from all components of the equation of motion. This means that we do not get any $\sim\K^3$ terms in the surface equation of motion using the LM method. No matter what the HEE functional for R$^2$ theory is, it is unlikely that no $\sim\K^3$ terms will occur in its surface equation of motion at any point in its parameter space. Even after taking the limit as prescribed, for general $R^2$ theory, there remain extra divergences in the bulk equation of motion. It is impossible to set these divergences to zero at all points of the parameter space, although this can be done for specific points like the Gauss-Bonnet point. 

As we discussed in the paper, the absence of $\sim\K^3$ terms is the R$^2$ equation of motion is an artifact of the way limits have to be taken in the LM method for the Gauss-Bonnet case. The limit can also be taken in such a way so as to preserve $\sim\K^3$ terms. It is worth recapitulating the results this way of taking the limit gives for Gauss-Bonnet theory. As we showed, using the second-order conical metric, the bulk equation of motion for Gauss-Bonnet theory, before taking the limit, has divergences only in the $zz$ and $ij$ components. There is no divergence in the $z\zb$ component, while the divergence in the $zi$ component turns out to be a constraint equation that vanishes by itself on using the Codazzi-Mainardi relation on AdS space. This same constraint equation results from the Jacobson-Myers functional, as well, on taking tangential variations of the surface. It is not clear what the relevance of the divergence in the $ij$ component is in the LM method. Were we to ignore this divergence, the surface equation of motion that would result from the $zz$ component for Gauss-Bonnet theory, after taking the limit, is $c\, \K =0$, where $c$ is proportional to the Weyl anomaly. This equation is clearly not what comes from the Jacobson-Myers functional. However, the resulting minimal surface is what one obtains on extremizing just the Wald entropy part of the functional. It would be interesting to check whether the $zz$ component of R$^2$ theory also leads to the same result. 

One of the pending issues with the LM method is to fix the ambiguity present in the limit-taking procedure. However, fixing this by itself does not seem enough to simultaneously cure the two problems present for R$^2$ theory: the absence of $\sim\K^3$ terms and the presence of extra constraints; although, it can remove one of these problems from the list. The ambiguity in the limit-taking procedure is not unique to the LM method. Similar, though not exactly the same, issues occur in studies of co-dimension two branes in the context of brane-world gravity \cite{Bostock}. It is possible that a further modification to the LM method will fix these problems; on the contrary, it may be that one cannot get rid of it in any way. The problem of extra divergences is related to the derivative order of the bulk equation of motion and seems to spring from the pathology of the general R$^2$ theory itself. In this sense, it is not surprising that we encounter it for general higher-derivative theories. Higher-derivative theories are known to suffer from problems regarding unitarity~\cite{Deser, Unitary, Ali}. These problems seem to be manifested in the LM method in the inability to remove all divergences, that occur on using the conical metric, from the bulk equation of motion.

What does our analysis say about the validity of the formula proposed in Refs.~\cite{camps,dong} as the entanglement entropy functional? For general R$^2$ theory as we demonstrated the leading-order terms match on both sides, which stops short of being a validation of the proposal for this theory. This test, at present, is similar in scope in refining conjectured entropy functionals for higher-derivative theories as the test whether the entropy functional leads to the correct universal terms. As we showed in this paper, for quasi-topological theory the universal terms are not sensitive to terms of the form  $\sim\K^4$ in the entropy functional (similar statement applies for other higher-derivative theories) and one can change these terms and still have the universal terms come out to be correct.  

The LM method, therefore, in its current form has limitations that make it ineffective in testing proposed entropy functionals for generic higher-derivative gravity theories.
The fact that the LM method only works for specific theories may indicate one of two things. One possibility is that 
entropy functionals only exist for specific theories such as Lovelock theories, for which the result of the surface equation of motion from the existing entropy functional and the LM method coincide. The other possibility, as mentioned before, is that the LM method needs some modification. In this context, it is also desirable that alternate methods to test entropy functionals be developed.


\section*{Acknowledgements}
We are grateful to Aninda Sinha for valuable suggestions and  remarks. We also thank Joan Camps and Rajesh Gopakumar for discussions. A.B thanks the string theory group at Harish-Chandra Research Institute, Allahabad for hospitality during part of this work. A.B also thanks the members of  the department of particle physics, University of Santiago de Compostela, specially Jose Edelstein, the Max Planck Institute of Gravitational Physics, Golm specially Axel Kleinschmidt and the theory division of Max Planck Institute for Physics, Munich specially Johanna Erdmenger for hospitality and the opportunity to present part of this work.

\appendix
\section{Conical Metric}\label{K2}
Near the conical singularity the bulk metric can also be written as  
\be
	ds^{2}= {\rho(x,y)}^{-2\epsilon} (dx^2+ dy^2) +  {\rho(x,y)}^{-2\epsilon} a_{pj}\,dx^{p}dx^{j} +g_{ij}\,dx^{i}dx^{j} \,.
\ee
The two-dimensional part is written in cartesian coordinates $x$ and $y$ and $\rho(x,y) = \sqrt{x^2+y^2}$. We have written the metric upto terms first order in $x(y)$. The co-dimension two surface ($ \Sigma$) is located at $x=0$ and $y=0$. The metric  $g_{ij}$ can be written down order by order in $x(y)$ after expanding around the surface $\Sigma$ as
\be 
	g_{ij}=h_{ij}+x \,  \partial_{x}  g_{ij}\big|_{\Sigma}+y   \,\partial_{y} g_{ij}\big|_{\Sigma}+\cdots.
\ee
The surface tensor $h_{ij}$ is independent of $x$ and $y$.  The variable $a_{pj}\sim O(x)$.

The extrinsic curvatures for the co-dimension two surface ($\Sigma$) are defined as 
\be \label{ext}
\K_{sij}= e^{\beta}_{j}\nabla_{i} n_{s\beta}\big|_{\Sigma}=e^{\beta}_{j} (\partial_{i} n_{s\beta} - {\hat \Gamma}^{\delta}_{\alpha \beta}e^{\alpha}_{i} n_{s \delta} )\big|_{\Sigma}\,.
\ee
Expanding the Christoffel in terms of the metric and using the fact that the first term $e^{\beta}_{j} \partial_{i} n_{s\beta} $ vanishes it follows that
\be 
  \K_{xij}=\frac{1}{2} \partial_{x}\, g_{ij}\big|_{\Sigma},\, \K_{y ij}=\frac{1}{2} \partial_{y} g_{ij}\big|_{\Sigma} \,.
\ee
We now make the simplifying assumption that the metric $g_{ij}$ is independent of the co-ordinate $y$. Under this assumption, the extrinsic curvature $\K_{y ij}$ vanishes as $\partial_{y} g_{ij}$ vanishes.

The complex coordinates $z$ and $\zb$ used in the metric in \equ{metric} are related to $x$ and $y$ as
\be
  z=x + i y,\,\bar{z}=x-i y\,.
\ee
In these coordinates the metric is
\be
	ds^{2}= e^{2\rho(z,\bar z)} (dz d \bar z )+g_{ij}dx^{i}dx^{j} + 2 e^{2\rho(z,\bar z)} \A_i (\zb dz-z d\zb) dy^i \,,
\ee
where 
\be 
	g_{ij}=h_{ij}+z  \, \K_{zij}+\bar z   \,\K_{\bar z ij}+\cdots.
\ee

The extrinsic curvatures in this coordinate system are related to $\K_{x ij}$ and $\K_{y ij}$ as
\be
\K_{zij}=\frac{\K_{x ij}+i\,\K_{y ij}}{2},\,\,\K_{\bar z ij} =\frac{\K_{x ij}-i\,\K_{y ij}}{2}\,.
\ee
Since $\K_{y ij}=0$ we have
\be
\K_{zij}=\K_{\bar zij} \,.
\ee
Similar considerations apply to the second-order quantities $\Q$.


\section{Curvature Tensors}\label{append}
In this appendix, we list components of the curvature tensors for the metric in \equ{metric2}, that do not appear in the main text. We retain only terms uptil zeroeth-order in $z,\zb$. 

The components of the Christoffels are
\begin{align}
{\Gamma^z}_{zz}&=-\frac{\epsilon}{z}\,,\quad{\Gamma^{\bar z}}_{\bar z\bar z}=-\frac{\epsilon}{\bar z}\,, \quad{\Gamma^z}_{ij}=-e^{-2 \rho(z,\bar z)}\,\,\K^{\bar z}_{ij}\,,\quad  {\Gamma^{\bar z}}_{ij}=-e^{-2 \rho(z,\bar z)}\,\,\K^{ z}_{ij}\,,\nonumber\\
{\Gamma^i}_{zj}&=\frac{1}{2}\,\K^{i}_{zj}\,,\,\,\,\, {\Gamma^i}_{\bar zj}=\frac{1}{2}\,\K^{i}_{\bar zj}\,,\,\,\,
\Gamma^{i}_{jk}=\frac{1}{2}g^{il}(\partial_{j}g_{lk}+\partial_{k}g_{lj}-\partial_{l}g_{jk})\,,\nonumber\\
{\Gamma^{z}}_{zi}&= -2\mathcal{A}_{i}\,,\quad \Gamma^{\bar z}_{\bar z i}=2\mathcal{A}_{i}\,. 
\end{align}

The components of the Riemann tensor are
\begin{align}
  R_{pqij} &=2 e^{2\rho(z,\bar z)}\hat\ve_{pq}\mathcal{F}_{ij}+(\K_{pjk} \K_{qi}^{k} - \K_{pik} \K_{qj}^{k}) \,,\nn\\
  R_{zizj}&= \tfrac{1}{4} \K_{zjk} \K_{zi}^{k} - \Q_{zzij}- \tfrac{\epsilon}{2 z}\K_{z ij} \,,\nn\\
R_{zi\bar zj}&=\tfrac{1}{2}e^{2\rho(z,\bar z)} \mathcal{F}_{ij}- 2e^{2 \rho(z,\bar z)} \A_i \A_j  + \tfrac{1}{4} \K_{zjk} \K_{\bar zi}^{k} -\tfrac{1}{2} \Q_{z\bar zij}\,,\nn\\
R_{pijk}&=\tfrac{1}{2}( \nabla_k \K_{pij} - \nabla_j \K_{pik}) \,,\nn\\
  R_{ikjl} &= \R_{ikjl} +\tfrac{1}{2} e^{-2\rho(z,\bar z)}  (\K_{zil} \K_{\bar zjk}+\K_{\bar zil} \K_{ zjk} - \K_{zij} \K_{\bar z kl}- \K_{\bar zij} \K_{ z kl}) \,,
\label{Riemanns2}
\end{align}
where $\mathcal{F}_{ij} \equiv \pa_i \A_j - \pa_j \A_i$. 

The components of the Ricci tensor are
\begin{align}
R_{zi}&=\tfrac{1}{2}( \nabla^{j}\K_{zji}-\nabla_{i}\K_{z})\,,\nn\\
R_{ z z}&=\tfrac{1}{4}\K_{zij}\K^{ij}_{z}-\tfrac{1}{2}Q_{zz}-\tfrac{\epsilon}{2 z}\K_{z} \,,\nn\\
R_{\bar z z}&=\tfrac{1}{4}\,\,\K_{\bar z ij} \,\,\K^{ ij}_{z}-\tfrac{1}{2} Q_{z\bar z} - 2e^{2\rho(z,\bar z)}(\A_{i}\A^{i}- 3 \O)\,,\nn\\
R_{ij}&=e^{-2\rho(z,\bar z)}\big( \K^{k}_{\bar z j}\K_{z ik}+\K^{k}_{ z j}\K_{\bar z ik}-\tfrac{1}{2}\K_{\bar z ij}\K_{z}-\tfrac{1}{2}\K_{ z ij}\K_{\bar z} -2 \Q_{z\bar zij} \big)+ \R_{ij} - 8 \A_{i} \A_{j}\,.
\end{align}
As in the main text, $\nabla$ used in the above equations is the Van der Waerden-Bortolotti covariant derivative~\cite{diffgeo} defined in \equ{defnabla}.

The Ricci scalar is
\be
R=\mathcal{R}+ 24 \Omega-16 \mathcal{A}_{i}\mathcal{A}^{i}- e^{-2\rho(z,\bar z)}\Big(\K_{z}\K_{\bar z}-3\K_{\bar zij}\K^{ij}_{z}+4\mathcal{Q}_{z\bar z}\Big)\,.
\ee


\end{document}